\documentclass[12pt]{article}

\usepackage{epsfig,amsmath,amssymb,verbatim,mathrsfs}

\numberwithin{equation}{section}

\def\beq{\begin{eqnarray}}
\def\eeq{\end{eqnarray}}
\def\bea{\begin{eqnarray}}
\def\eea{\end{eqnarray}}

\def\gev{\, {\rm GeV}}

\newcommand{\gsim}{\lower.7ex\hbox{$\;\stackrel{\textstyle>}{\sim}\;$}}
\newcommand{\lsim}{\lower.7ex\hbox{$\;\stackrel{\textstyle<}{\sim}\;$}}

\def\stilde{\widetilde}

\newcommand{\newc}{\newcommand}
\newc{\Nc}{N_{c}}
\newc{\CG}{C_G}
\newc{\gp}{g'}
\newc{\stopi}{\stilde t_i}
\newc{\sboti}{\stilde b_i}
\newc{\staui}{\stilde \tau_i}
\newc{\stopj}{\stilde t_j}
\newc{\sbotj}{\stilde b_j}
\newc{\stauj}{\stilde \tau_j}
\newc{\stopI}{\stilde t_1}
\newc{\stopII}{\stilde t_2}
\newc{\sbotI}{\stilde b_1}
\newc{\sbotII}{\stilde b_2}
\newc{\stauI}{\stilde \tau_1}
\newc{\stauII}{\stilde \tau_2}
\newc{\sstop}{s_{t}}
\newc{\cstop}{c_{t}}
\newc{\ssbot}{s_{b}}
\newc{\csbot}{c_{b}}
\newc{\sstau}{s_{\tau}}
\newc{\cstau}{c_{\tau}}
\newc{\Sstop}{s_{2t}}
\newc{\Cstop}{c_{2t}}
\newc{\Ssbot}{s_{2b}}
\newc{\Csbot}{c_{2b}}
\newc{\Sstau}{s_{2\tau}}
\newc{\Cstau}{c_{2\tau}}
\newc{\salpha}{s_\alpha}
\newc{\calpha}{c_\alpha}
\newc{\Calpha}{c_{2\alpha}}
\newc{\Salpha}{s_{2\alpha}}
\newc{\sbetapm}{s_{\beta_\pm}}
\newc{\cbetapm}{c_{\beta_\pm}}
\newc{\Sbetapm}{s_{2 \beta_\pm}}
\newc{\Cbetapm}{c_{2 \beta_\pm}}
\newc{\sbetaO}{s_{\beta_0}}
\newc{\cbetaO}{c_{\beta_0}}
\newc{\SbetaO}{s_{2 \beta_0}}
\newc{\CbetaO}{c_{2 \beta_0}}
\newc{\vu}{v_u}
\newc{\vd}{v_d}
\newc{\seL}{\stilde e_L}
\newc{\smuL}{\stilde \mu_L}
\newc{\seR}{\stilde e_R}
\newc{\smuR}{\stilde \mu_R}
\newc{\suL}{\stilde u_L}
\newc{\sdL}{\stilde d_L}
\newc{\suR}{\stilde u_R}
\newc{\sdR}{\stilde d_R}
\newc{\scL}{\stilde c_L}
\newc{\ssL}{\stilde s_L}
\newc{\scR}{\stilde c_R}
\newc{\ssR}{\stilde s_R}
\newc{\snue}{\stilde \nu_e}
\newc{\snumu}{\stilde \nu_\mu}
\newc{\snutau}{\stilde \nu_\tau}
\newc{\Gpm}{G^\pm}
\newc{\Hpm}{H^\pm}
\newc{\FFbS}{\overline{FF}S}
\newc{\FFbV}{\overline{FF}V}
\newc{\FSS}{F_{SS}}
\newc{\FSSS}{F_{SSS}}
\newc{\FFFS}{F_{FFS}}
\newc{\FFFbS}{F_{\overline{FF}S}}
\newc{\FSSV}{F_{SSV}}
\newc{\FVS}{F_{VS}}
\newc{\FVVS}{F_{VVS}}
\newc{\FFFV}{F_{FFV}}
\newc{\FFFbV}{F_{\overline{FF}V}}
\newc{\Fgauge}{F_{\rm gauge}}
\newc{\DRbarprime}{$\overline{\rm DR}'$ }
\newc{\DRbar}{$\overline{\rm DR}$ }
\newc{\MSbar}{$\overline{\rm MS}$ }
\newc{\Yu}{{\bf Y}_u}
\newc{\Yd}{{\bf Y}_d}
\newc{\Ye}{{\bf Y}_e}
\newc{\Au}{{\bf a}_u}
\newc{\Ad}{{\bf a}_d}
\newc{\Ae}{{\bf a}_e}
\newc{\bm}{{\bf m}}
\newc{\zhol}{Z^{\rm hol}}

\newcommand{\lal}{\lambda A_{\lambda}}

\newcommand{\kol}{\frac{\kappa}{\lambda}}
\newcommand{\kolsq}{\frac{\kappa^2}{\lambda^2}}
\newcommand{\stb}{s_{2\beta}}
\newcommand{\ctb}{c_{2\beta}}
\newcommand{\ccdot}{\!\cdot\!}
\newcommand{\nnmb}{\nonumber}

\newcommand{\lrf}[2]{\left(\frac{#1}{#2}\right)}
\newcommand{\dash}{\!-\!}
\newcommand{\ffb}{\textbf{5}\oplus\bar{\textbf{5}}}
\newcommand{\alfer}{\frac{\lambda}{\kappa}\frac{A_{\lambda}}{\mu}}

\newcommand{\ltk}{\tilde{\lambda}}

\begin{document}

\setlength{\baselineskip}{0.2in}



\begin{titlepage}
\noindent
\begin{flushright}
MCTP-08-55\\
\end{flushright}
\vspace{1cm}

\begin{center}
  \begin{Large}
    \begin{bf}
Modified Higgs Boson Phenomenology from Gauge or Gaugino Mediation
in the NMSSM\\
     \end{bf}
  \end{Large}
\end{center}
\vspace{0.2cm}

\begin{center}

\begin{large}
David E. Morrissey and Aaron Pierce\\
\end{large}
\vspace{0.3cm}
  \begin{it}
Michigan Center for Theoretical Physics (MCTP) \\
Physics Department, University of Michigan, Ann Arbor, MI 48109
\vspace{0.5cm}
\end{it}\\

\end{center}

\center{\today}

\begin{abstract}
In the Next-to-Minimal Supersymmetric Standard Model (NMSSM), 
the presence of light pseudoscalars can have a dramatic effect 
on the decays of the Standard Model-like Higgs boson.  These pseudoscalars 
are naturally light if supersymmetry breaking preserves an approximate 
$U(1)_{R}$ symmetry, spontaneously broken when the Higgs 
bosons take on their expectation values. We investigate two 
classes of theories 
that possess such an approximate $U(1)_{R}$ at the mediation scale:
deformations of gauge and gaugino mediation.  In the models 
we consider, we find two disjoint classes of phenomenologically allowed
parameter regions.  One of these regions corresponds to a
limit where the singlet of the NMSSM largely decouples. The other can give
rise to a Standard Model-like Higgs boson with a dominant
branching into light pseudoscalars.  
\end{abstract}

\vspace{1cm}

\end{titlepage}

\setcounter{page}{2}


\vfill\eject



\newpage

\section{Introduction}

  The Minimal Supersymmetric Standard Model (MSSM) 
is a leading theory for new physics at the weak scale, 
providing for a stabilization of the weak hierarchy, 
the unification of gauge couplings, and an excellent dark matter 
candidate~\cite{Martin:1997ns}.  In contrast to the 
Standard Model~(SM), the MSSM possesses a pair 
of electroweak Higgs doublets. These give rise to five 
physical degrees of freedom instead of the single SM  Higgs boson.
Despite this complication, throughout much of the allowed
MSSM parameter space one of the MSSM Higgs bosons behaves in very
much the same way as the SM Higgs~\cite{Carena:2002es}.
  
  The MSSM is not without its puzzles.  In particular, 
it possesses a dimensionful parameter in the superpotential that 
marries the two Higgs multiplets together:  $W \supset \mu H_{u} H_{d}$.  
This parameter must be of order the weak scale in order to 
achieve proper electroweak symmetry breaking.  
One approach to this problem is the Next-to-Minimal Supersymmetric 
Standard Model~(NMSSM), which consists of the MSSM augmented by a singlet 
chiral superfield~\cite{Frere:1983ag,Ellis:1988er}.  
This singlet superfield leads to two new physical 
Higgs states: a neutral scalar and 
a neutral pseudoscalar.  The appearance of these
states, along with their mixing with the MSSM Higgs states,
can significantly modify the Higgs phenomenology at the
 Large Hadron Collider (LHC)
and other colliders~\cite{Ellis:1988er,Ellwanger:1993xa,
Ellwanger:1999ji,Miller:2003ay,Barger:2006dh}.

  The LHC Higgs boson signatures of the NMSSM are particularly different 
from the SM (and the MSSM) when the light SM-like Higgs boson can decay
into pairs of very light mostly singlet Higgs pseudoscalars.  
(For a recent review of these decays see Ref.~\cite{Chang:2008cw}.)  
It was observed in Ref.~\cite{Dermisek:2005ar} that if this decay 
mode dominates over the standard SM modes such as $h\to b\bar{b}$, 
the LEP-II bound on the SM-like Higgs mass is lowered below $114\,\gev$.
Masses of $110\,\gev$ or below are possible if the pseudoscalars 
decay primarily into bottom quarks~\cite{Chang:2008cw,Schael:2006cr}, 
and Higgs bosons as light as $90\,\gev$ can be consistent with the LEP 
data if the pseudoscalars decay primarily into tau leptons.  
Since the dominant contribution to fine-tuning within the MSSM 
comes from the Higgs boson mass bound, reducing the 
bound on the SM-like Higgs mass in this way can conceivably ameliorate
the MSSM Higgs-sector fine-tuning problem.  Recent discussions of this 
point have appeared in Refs.~\cite{
Dermisek:2005ar,Chang:2005ht,Schuster:2005py,Dermisek:2006wr}.  

  Often, the Higgs pseudoscalars of the NMSSM 
are too heavy for the SM-like Higgs to decay into them. 
An important exception occurs when the theory has an 
approximate continuous symmetry 
under which at least one of the Higgs states is charged.  
In this case, a light pseudoscalar arises as the 
pseudo-Nambu-Goldstone boson (pNGB) of the approximate
symmetry when it is broken in the course of electroweak symmetry breaking.  
A promising candidate for this symmetry is a  
$U(1)_R$~\cite{Schuster:2005py,Dermisek:2006wr,Dobrescu:2000jt}, 
which is exact in the limit where the trilinear $A$ terms of 
the Higgs sector vanish.  

This symmetry is realized approximately      
in a natural way if the mediation of supersymmetry breaking 
is dominated by gauge interactions, and the Higgs sector $A$ terms vanish 
at the mediation scale.  In this
case the dominant source 
of $U(1)_R$ violation comes from the (Majorana) gaugino masses.  
This breaking is communicated to the Higgs fields in the course 
of renormalization group~(RG) running between the mediation scale and 
the electroweak scale.  

 Two ways to mediate supersymmetry breaking through gauge interactions with
 vanishing $A$ terms are gauge mediation~(GMSB)~\cite{Dine:1982zb,Dine:1993yw,
Giudice:1998bp}
and gaugino mediation~(\~gMSB)~\cite{Kaplan:1999ac,Chacko:1999mi}.  
In both of these mechanisms of SUSY breaking, it is challenging to 
generate both the 
$\mu$ and the $B\mu$ terms with the correct relative size in the MSSM. 
One way to address this challenge is to add a singlet superfield 
whose vacuum expectation value (VEV) 
generates these terms, as is done in the NMSSM.  
Unfortunately, previous work on combining gauge or gaugino mediation
with the NMSSM has found that the effective shielding of the
singlet sector from the supersymmetry breaking also makes it
difficult to obtain an acceptable pattern of electroweak
symmetry breaking~\cite{deGouvea:1997cx}.  A number of works have 
investigated extensions of minimal gauge mediation in the hope
of relieving this tension~\cite{Agashe:1997kn,Delgado:2007rz,
Giudice:2007ca,Liu:2008pa}. 

  In the present work, we study the LHC Higgs boson phenomenology of two  
simple extensions of minimal gauge and gaugino mediation within
the NMSSM.  The first extension consists of relaxing 
some of the GMSB boundary conditions on the singlet-sector soft terms. 
This might arise from a direct coupling of the singlet sector 
to the messenger states, or if there is additional source of 
supersymmetry breaking coupled solely to the singlet \cite{Djouadi:2008uw}. 
Such a modification of gauge mediation was considered recently
in Ref.~\cite{Ellwanger:2008py} while this work was in preparation.  
We extend and expand upon their results
within gauge mediation, and apply the same modification to gaugino mediation.
The second extension we consider involves adding new charged 
vector-like states to the theory~\cite{Agashe:1997kn,deGouvea:1997cx}.  
These modify the low-energy soft parameters of the NMSSM through their effects 
on the renormalization group~(RG) running.  
Before discussing these modifications in detail, we introduce notation 
and review why a modification of the minimal gauge mediation 
scenario is needed.

\subsection{The NMSSM Higgs Sector}

  The Higgs sector of the NMSSM consists of the $H_u$ and $H_d$
doublets of the MSSM along with a singlet $S$.  The corresponding 
superpotential is given by\footnote{
We follow the notation and sign conventions 
of NMHDECAY~\cite{Ellwanger:2004xm}.}
\beq
W \supset 
\lambda\,S H_u\ccdot H_d + \frac{1}{3}\kappa\,S^3 , 
\eeq
where $A\ccdot B := \epsilon^{ab}\,A_a B_b$ with $\epsilon^{12} = +1$.
The soft supersymmetry breaking operators within the Higgs sector are 
\beq
-\mathscr{L}_{soft} &\supset& m_{H_u}^2|H_u|^2 + m_{H_d}^2|H_d|^2 
+ m_S^2|S|^2 + \left( \lambda A_{\lambda} H_u\ccdot H_d S 
+ \frac{1}{3}\kappa A_{\kappa}S^3+h.c.\right) 
\eeq
The NMSSM Lagrangian as shown has a $\mathbb{Z}_3$ discrete symmetry
that forbids the appearance of linear and quadratic singlet operators.
When the singlet obtains a VEV in the early universe, this symmetry is 
broken spontaneously in one of three degenerate vacua, and dangerous 
domain walls can form~\cite{Abel:1995wk}.  
These can be avoided by including a relevant operator that softly
breaks the $\mathbb{Z}_3$ well below the electroweak scale.  
Such operators can arise in a natural way~\cite{Abel:1996cr,
Panagiotakopoulos:1998yw} 
such that they eliminate domain walls, but are too small to have 
a significant effect on the Higgs boson phenomenology.  

  It is useful to introduce a complete basis for the set of all 
$U(1)$ transformations on $S$, $H_u$, $H_d$. 
\beq
\begin{array}{c|cccccccc}
&S&H_u&H_d&Q&U^c&D^c&L&E^c\\
\hline
{\bf PQ}&-2&1&1&-1&0&0&-1&0\\
{\bf R}&2/3&2/3&2/3&2/3&2/3&2/3&2/3&2/3\\
{\bf PQ'}&0&1&-2&0&-1&2&0&-1
\end{array}
\eeq
All of these couplings are broken by terms in the 
superpotential or the soft SUSY-breaking Lagrangian. 
$U(1)_{PQ}$ is broken explicitly by $\kappa$ and $\kappa A_{\kappa}$,
$U(1)_{R}$ is broken by $\lambda A_{\lambda}$,
$\kappa A_{\kappa}$, and the other trilinear soft terms and gaugino
masses, while  $U(1)'_{PQ}$ is broken explicitly by $\lambda$
and $\lal$.  Throughout this paper, we will implicitly make a $U(1)_{PQ}'$
field redefinition such that $\lambda$ is real and positive,
as well as a $U(1)_{PQ}$ transformation such that $\kappa$ is real.
With a  $U(1)_R$ rotation, we can take gaugino masses to all be 
real and positive provided they have no relative phases.  
The $A$ terms subsequently generated by RG running will then be 
(mostly) real as well.  
Having fixed $\kappa$ to be real, but not necessarily positive, 
there is still a residual $\mathbb{Z}_2$ subgroup of $U(1)_{PQ}$ 
that can be used (in conjunction with $U(1)_Y$), 
to make the VEVs $\left<H_u^0\right> = v_u$ and $\left<H_d^0\right> =v_d$ 
real and positive.  The values of $\left<S\right> = v_s$ 
and $\kappa$ can then take either sign.

\subsection{Minimization Conditions and a Challenge}

  The Higgs-sector parameters in the NMSSM are
$\{ \lambda,\,\kappa,\,A_{\lambda},\,A_{\kappa},\,
m_{H_u}^2,\,m_{H_d}^2,m_{S}^2\}$.  Electroweak symmetry
breaking provides three minimization conditions that allow
three of these parameters to be exchanged for three VEVs.
While it is a common procedure to solve for the soft masses 
$m_{H_u}^2$ and $m_{H_d}^2$ in terms of $v$ and $\tan\beta$, 
we find that it is instructive to instead solve for 
$m_S^2$, $\mu = \mu_{eff} = \lambda\,v_s$, 
and $\kappa$ in terms of the other parameters.  

  Demanding that $v_u$, $v_d$, and $v_s$ are all non-zero and real,
we find
\bea
\mu &=& sgn(\mu)\,\sqrt{ \frac{m_{H_d}^2-m_{H_u}^2\tan^2\beta}{\tan^2\beta-1}
-\frac{1}{2}M_Z^2\phantom{a}},
\label{musoln}\\
\frac{\kappa}{\lambda} &=& \frac{\sin2\beta}{2}\,\left(
2+\frac{\lambda^2v^2}{\mu^2} + \frac{m_{H_u}^2+m_{H_d}^2}{\mu^2} \right)
-\frac{A_{\lambda}}{\mu},
\label{kapsoln}\\
\frac{m_S^2}{\mu^2} &=& 
-2\kolsq
+\frac{\sin2\beta}{2}\,(\frac{A_{\lambda}}{\mu}+\kol)\,
\frac{\lambda^2v^2}{\mu^2}
-\kol\,\frac{A_{\kappa}}{\mu}.
\label{mssoln}
\eea
These relations imply a difficulty for the NMSSM in 
gauge mediation~\cite{deGouvea:1997cx}.  
We now review this argument.  
To leading order, the trilinear singlet soft terms 
and $m_S^2$ vanish at the messenger 
scale and are generated primarily from RG 
running to the electroweak scale. 
For lower mediation scales, the logarithm does not compensate the loop 
factor suppression, and these terms are expected to be somewhat 
smaller than the other soft terms.  On the other hand under the condition that 
the lightest slepton is sufficiently heavy, the soft 
terms generated by minimal GMSB force $|m_{H_u}^2|$ to be larger 
than about $(250\,\gev)^2$.
Eq.~\eqref{musoln} then forces $\mu^2$ to be about the same 
size as $|m_{H_u}^2|$ (for moderate $\tan\beta$).
These large values of $\mu^2$ lead to the hierarchies
$\lambda^2\,v^2/\mu^2 \ll 1$ and $|A_{\lambda,\kappa}/\mu| \ll 1$.
To satisfy Eq.~\eqref{mssoln} given the small values of $m_S^2/\mu^2$ 
expected from minimal gauge mediation, it is necessary to have 
$|\kappa/\lambda| \ll 1$.  As a result, $\sin2\beta \ll 1$ 
is needed for Eq.~\eqref{kapsoln} to be satisfied.  

  With both $|\kappa/\lambda|,\,\sin2\beta\ll 1$, it is difficult 
to obtain an acceptable pattern of symmetry breaking.  
In this limit, the phenomenologically interesting local 
extremum (corresponding to the solutions of
Eqs.~(\ref{musoln},\ref{kapsoln},\ref{mssoln}) where $v_u$, $v_d$, 
and $v_s$ are all non-zero) tends to be a saddle
point rather than a minimum.  This can be seen in 
the determinant of the $CP$-even Higgs boson mass-squared matrix,
\beq 
\det\mathcal{M}_S^2 \simeq
4\mu^4\,\lambda^2v^2\,\kol\,\frac{2}{\sin2\beta}\,
\left(\kolsq\frac{{g'}^2+g^2}{2\lambda^2} -1\right).
\label{msdet}
\eeq
With $|\kappa/\lambda| \ll 1$, this expression is negative 
unless $\lambda^2\ll 1$ is also tiny.
Very small values of $\lambda$ and $\kappa$, with $\mu$ fixed, 
correspond to a decoupling limit of the NMSSM
in which the singlet states couple only very weakly 
to the rest of the theory, and the Higgs phenomenology
reduces to that of the MSSM.
The argument presented here applies to minimal gaugino 
mediation as well.

  Eqs.~(\ref{musoln}-\ref{mssoln}) also indicate a way out 
of this difficulty: a deformation of minimal gauge or gaugino mediation 
that generates a large negative value of $m_S^2$ near the electroweak scale. 
In the present work, we investigate the phenomenology of two of the
simplest possibilities for modifying the low-scale value of $m_S^2$ 
in gauge and gaugino mediation.
The first modification we consider consists of treating
$m_S^2$ as a free parameter at the mediation scale of
supersymmetry breaking.  The second consists of adding 
charged vector-like states to the theory with superpotential 
couplings to the singlet~\cite{Dine:1993yw,Agashe:1997kn,deGouvea:1997cx}.
These couplings drive the singlet soft mass negative in
the course of RG running, in a manner analogous to the 
effect of the top Yukawa coupling in the MSSM.  The goal of the 
present work is to determine the viability of these modifications and to 
investigate the LHC Higgs signatures they predict.  
We do not dwell on a reduction of fine-tuning relative to the MSSM.  
Rather, our focus is to better understand under what conditions 
one might expect novel Higgs boson signatures.  Then, if such signatures 
are indeed observed at colliders, we will have an important clue 
about the identity of the underlying theory.

  The plan of the remainder of this paper is as follows.  
In Section~\ref{u1r} we derive analytic expressions for
the Higgs boson masses and couplings in the approximate $U(1)_R$ limit.
In Section~\ref{free} we study the NMSSM Higgs sector within 
gauge mediation with an additional
boundary contribution to $m_S^2$ at the gauge messenger scale.
Next, in Section~\ref{gauge} we investigate the effect of adding
charged vector-like states on the NMSSM Higgs states 
within minimal (unmodified) gauge mediation.  We study analogous
deformations of gaugino mediation within the NMSSM in Section~\ref{ino}.
Section~\ref{concl} is reserved for our conclusions.  Some of our
technical results are collected in Appendices~\ref{appa} and \ref{appb}.

\section{NMSSM Higgs Bosons with an Approximate $U(1)_R$\label{u1r}}

  We begin by deriving approximate analytic expressions 
for the Higgs boson masses, mixing angles, and couplings in the NMSSM 
when the theory has an approximate $U(1)_R$ symmetry in 
the singlet sector~\cite{Dobrescu:2000jt}.  This approximate 
symmetry is realized numerically as the hierarchies 
$|A_{\lambda}/\mu|\ll 1$ and $|A_{\kappa}/\mu| \ll 1$.
We shall also assume $|\lambda\,v/\mu| \ll 1$, which is valid throughout 
most of the parameter space of the models 
we study in the coming sections.  Further details about this
expansion are listed in Appendix~\ref{appa}.  

  To leading non-trivial order in the small ratios,
and assuming $\kappa/\lambda$ and $\sin 2\beta$ are
not too terribly small, the tree-level $CP$-odd masses are 
\bea
m_{a_s}^2 &=& 3\frac{\kappa}{\lambda}\mu\,\left(
\frac{3\lambda}{\kappa}\frac{\stb}{2}
\frac{\lambda^2v^2}{\mu^2}\,{A_{\lambda}} 
- A_{\kappa}\right),
\label{cpodd1}\\
m_{A^0}^2 &=&
\left(1+\frac{\lambda}{\kappa}\frac{A_{\lambda}}{\mu}\right)
\frac{2}{\stb}\frac{\lambda}{\kappa}\left(\frac{\kappa}{\lambda}\mu\right)^2
+ 2\frac{\kappa}{\lambda}\stb
\left(1-2\frac{\lambda}{\kappa}\frac{A_{\lambda}}{\mu}\right)\lambda^2v^2.
\label{cpodd2}
\eea
The expression for $m_{a_s}^{2}$ vanishes in the limit 
$A_{\lambda,\kappa}\to 0$
showing that this state is the pNGB of the approximate $U(1)_R$.
This state is primarily singlet, while the $A^0$ state is similar 
to the Higgs pseudoscalar in the MSSM.  We are able to make this 
identification because the mixing among the singlet and 
the non-singlet pseudoscalars is suppressed by a factor of 
$\lambda\,v/\mu \ll 1$.  Full expressions for the mass and 
mixing matrices in this expansion are listed in Appendix~\ref{appa}.

  Applying this expansion to the $CP$-even masses yields
\bea
m_{h^0}^2 &=& \lambda^2v^2\left[\frac{{g'}^2+g^2}{2\lambda^2}\ctb^2
+\stb^2 - \left(\frac{\lambda}{\kappa}-\stb\right)^2
\right],
\label{cpeven1}\\
m_{H^0}^2 &=& \frac{2}{\stb}
\frac{\lambda}{\kappa}\,\left(\frac{\kappa}{\lambda}\mu\right)^2, 
\label{cpeven2}\\
m_{h_s}^2 &=& 4\left(\frac{\kappa}{\lambda}\mu\right)^2. 
\label{cpeven3}
\eea
Among these states, $h^0$ is SM-like, $H^0$ is similar to the
corresponding state in the MSSM, and $h_s$ is predominantly
singlet. Again, this identification is possible because the
mixing between the MSSM states and the singlet is suppressed
by factors of $\lambda\,v/\mu \ll 1$.  The $CP$-even sector mixing 
matrices are also listed in Appendix~\ref{appa}.

  The (exact) mass of the charged Higgs is
\beq
m_{H^{\pm}}^2 = \frac{2}{\stb}\frac{\lambda}{\kappa}\left(\kol\mu\right)^2
\left(1+\frac{\lambda}{\kappa}A_{\lambda}\right) 
+ \lambda^2\,v^2\,\left(\frac{g^2}{2\lambda^2}-1\right).
\eeq
In the limit of $\mu^2 \gg \lambda^2v^2,\, A_{\lambda}^2,\;A_{\kappa}^2$
this coincides closely with the masses of the $A^0$ and $H^0$ states.

  The coupling between the SM-like $h^0$ state and pairs of the 
light $a_s$ pseudoscalars is particularly important for the phenomenology 
of this scenario.  This coupling corresponds to the operator 
\beq
\mathcal{L} \supset \frac{c}{\sqrt{2}}\,v\,h^0\,a_1\,a_1.
\eeq
Within the expansion, the coefficient $c$ is given by
\bea
c &=& \left(\frac{1}{2}\lambda^2\right)\,
\left(\frac{\lambda}{\kappa}-2\,\stb\right)\,
\left(\frac{\lambda}{\kappa}+\stb\right)\,
\frac{m_{h^0}^2}{2\mu^2}\label{cform}\\
&&
+\left(\frac{1}{2}\lambda^2\right)\left[
\frac{1}{2}\frac{\lambda}{\kappa}\frac{A_{\kappa}}{\mu}
\left(1-\frac{\kappa}{\lambda}\stb
-12\,\frac{\kappa^2}{\lambda^2}\stb^2\right)
-9\left(\frac{\kappa}{\lambda}\stb\right)
\frac{\lambda}{\kappa}\frac{A_{\lambda}}{\mu}
\left(1-\frac{\lambda}{\kappa}\frac{A_{\lambda}}{\mu}\right)\right].
\nnmb
\eea
The first line in this expression coincides with the result of 
Ref.~\cite{Dobrescu:2000jt}, and corresponds to the axion-like derivative 
coupling of the light pNGB pseudoscalar.  The second line is new to 
our calculation.  It is useful because it captures the contributions 
to the coupling that arise from the explicit breaking of the $U(1)_R$ 
symmetry by the $A$-terms.  
In terms of the coefficient $c$, the decay width for $h^0\to a_sa_s$ is
\beq
\Gamma(h^0\to a_sa_s) = \frac{c^2v^2}{16\pi\,m_{h^0}}
\left(1-4\frac{m_{a_s}^2}{m_{h^0}^2}\right)^{1/2}.
\eeq
The approximate expression in Eq.~\eqref{cform} agrees well with the
full numerical result from NMHDECAY\cite{Ellwanger:2004xm} in 
the appropriate limit.

\section{Gauge Mediation in the NMSSM with a Free $m_s^2$\label{free}}

  The first scenario we consider is a deformation 
of minimal gauge mediation in which the value of the singlet 
soft mass $m_S^2$ taken to be a free parameter at the 
messenger scale $M$.  All other soft terms
are set to their standard gauge mediated values at scale $M$.
Without this deformation, $m_S^2(M)$ vanishes at the leading order.
This feature is the primary obstruction to merging gauge 
mediation with the NMSSM~\cite{deGouvea:1997cx}.  
By liberating $m_S^2$ from its minimal GMSB 
boundary condition we avoid this obstacle by fiat.
The same deformation was considered in Ref.~\cite{Ellwanger:2008py}. 
We expand upon and extend their results.

  We do not have a particular model in mind for how this deformation
of minimal gauge mediation could arise.  However, as a gauge singlet, 
one might imagine that $S$ is on a special footing and might feel 
the mediation of supersymmetry differently from the 
rest of the MSSM.\footnote{This philosophy is similar 
in spirit to \cite{Baer:2004fu,HENS}, where supersymmetry boundary conditions 
for the Higgs fields were chosen to be different from the matter fields.} 
A number of recent works have considered 
related modifications of gauge mediation within
the NMSSM with this fact in 
mind~\cite{Delgado:2007rz,Liu:2008pa,Ellwanger:2008py}. 
Prior explicit constructions typically generate new contributions 
to $A_{\lambda}$ and $A_{\kappa}$.  This 
spoils the approximate $U(1)_R$ symmetry, and prevents the 
interesting decays to pseudoscalars.  Modifying $m_S^2$ without 
altering $A_{\lambda}$ and $A_{\kappa}$ requires coupling $S$ directly
to a source that breaks supersymmetry but not the $U(1)_R$.

  We do not explicitly write a full model of supersymmetry breaking 
for the singlet field.  However, we point out that supersymmetry 
breaking without spontaneous $U(1)_R$ breaking is a generic feature
of simple O'Raifertaigh models~\cite{Shih:2007av}, 
and that some care and complication is often required to break 
the $U(1)_R$ in this context~\cite{Shih:2007av,Intriligator:2007py}. 
It is a model-building challenge to ensure that the additional 
SUSY breaking felt by the singlet field is of the same order as 
the rest of the SUSY breaking.

\subsection{Allowed Parameter Regions}

  We begin by searching for phenomenologically acceptable 
regions of the NMSSM parameter space subject to GMSB boundary 
conditions for all soft terms other than $m_S^2$.
We specify the superpotential coupling $\lambda$, as well as the  
supersymmetry breaking scale $F/M$, at the GMSB messenger scale, $M$. 
We specify $\tan\beta$ near the electroweak scale.  With these inputs set, 
we compute the resulting low-energy spectrum.  
The soft supersymmetry breaking parameters at scale $M$,
other than $m_S^2$, are set to their gauge-mediated values assuming 
a minimal GMSB sector with a single set of $\ffb$ messengers.  
We use a modified version of NMSPEC/NMHDECAY~\cite{Ellwanger:2004xm} 
to perform the RG evolution of the model parameters, to find the low-scale 
values of $\kappa$, $|\mu|$, and $m_S^2$, and to compute
the low-energy spectrum and constraints.  In general, the value of 
$m_S^2$ obtained in this way does not agree with the GMSB boundary 
condition of $m_S^2(M) \simeq 0$.  Our scan encompasses the 
parameter ranges
\beq
\lambda\in [0.001,0.7],~~\tan\beta\in [1,50],~~
M \in [10^5,10^{14}]\gev,~~F/M \in [2,40]\times 10^4\,\gev.
\eeq
These ranges lead to low-scale gluino masses between about
$350\,\gev$ and $2500\,\gev$.

\begin{figure}[ttt]
\vspace{1cm}
\begin{center}
        \includegraphics[width = 0.5\textwidth]{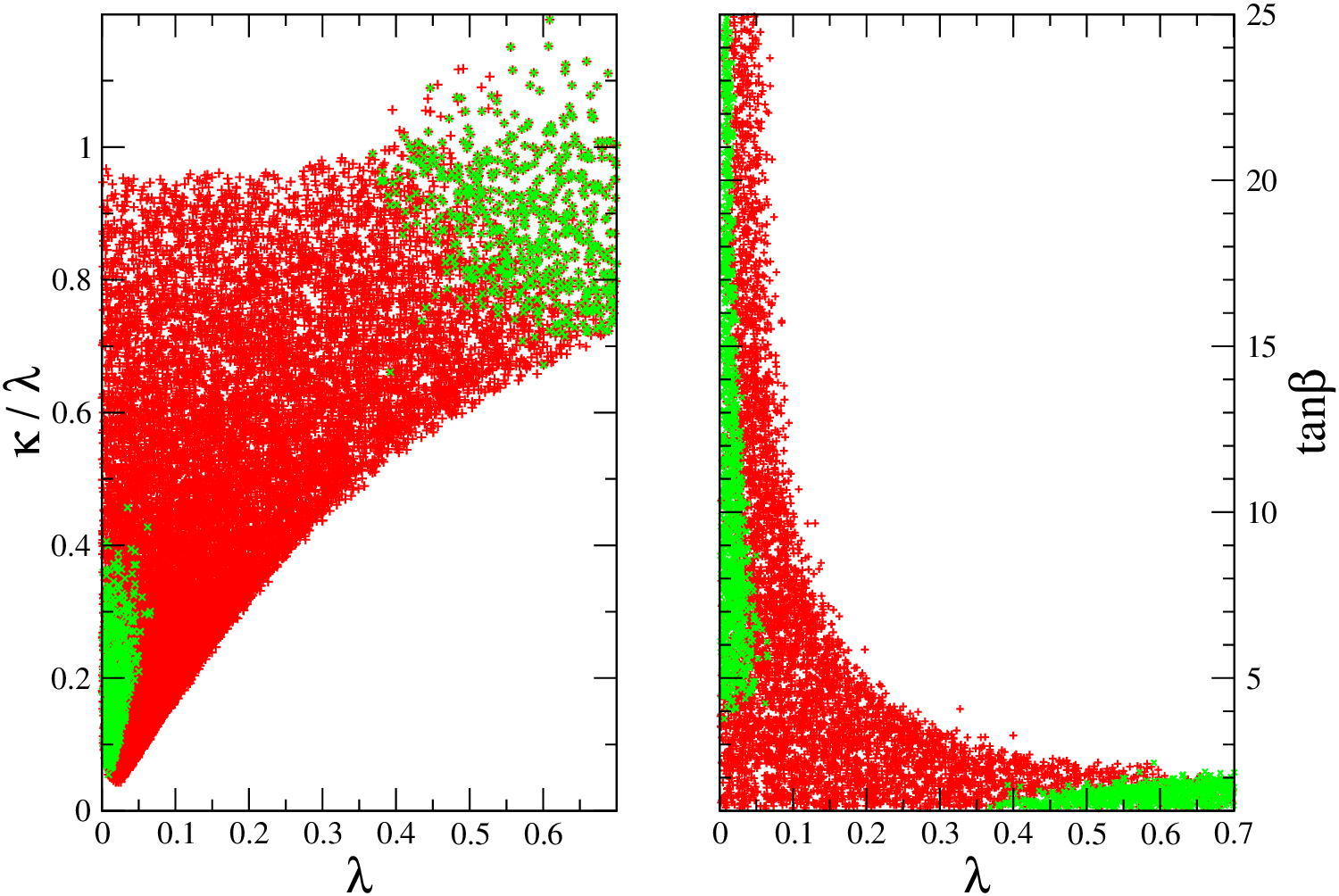}
\end{center}
\caption{Allowed parameter points for the NMSSM with 
minimal GMSB boundary conditions for a single set of $\ffb$ messengers,
but with no restrictions on $m_S^2$.  The red points do not include 
Higgs constraints while the green points do.
}
\label{gauge-ms-lk}
\end{figure}

  The results of our parameter scans are shown in Fig.~\ref{gauge-ms-lk}
within the $\lambda\dash\kol$ and $\lambda\dash\tan\beta$ planes.
The red (dark) points in these plots are consistent with all relevant 
phenomenological bounds except for the collider constraints 
on the neutral $CP$-even and $CP$-odd Higgs bosons, 
while the green (light) points satisfy the Higgs constraints as well.  
We do not demand that the lightest neutralino be the 
LSP since for gauge mediation the gravitino is usually expected 
to be the true LSP.  However, we do require that the couplings 
remain perturbative up to the scale $M$, but not $M_{GUT}$.  
This eliminates points with large $\lambda$ and $\kappa$ for 
larger mediation scales.  In the limit $M \to M_{GUT}$, we find that
perturbativity requires $(\lambda^2+\kappa^2) \lesssim 0.45$
near the electroweak scale~\cite{Miller:2003ay,Liu:2008pa}.  
When $M < M_{GUT}$, additional charged states that enter the 
running at $M$, such as the messengers themselves, can help to 
slow down the growth of $\lambda$ and $\kappa$ above 
the messenger scale~\cite{Barbieri:2007tu}.

  In the left panel of Fig.~\ref{gauge-ms-lk},
the lower limit on the red (dark) region corresponds
to the condition $\det\mathcal{M}_S^2>0$.  For smaller values of
$\lambda$, this cutoff agrees with the relation given 
in Eq.~\eqref{msdet}, while more generally, it coincides with
$m_{h^0}^2 > 0$ using the expression in Eq.~\eqref{cpeven1}.
The upper boundary of the allowed red (dark) region in Fig.~\ref{gauge-ms-lk}
can be understood from an examination of Eq.~\eqref{kapsoln}. 
There is a close relationship between $\kappa/\lambda$ 
and $\sin2\beta$ when $|\mu|$ is much larger than $\lambda\,v$ 
and the singlet trilinear couplings.  Since $\sin 2 \beta$ 
is necessarily bounded by $1$, so is $\kappa/ \lambda$.

  It is clear from Fig.~\ref{gauge-ms-lk} that there are two disjoint 
populations of phenomenologically consistent (light/green) parameter points.  
The first population, which we call Region~I, has larger values of
$\lambda \gtrsim 0.4$, $\kappa/\lambda$ on the order of unity, 
and smaller values of $\tan\beta\lesssim 2.5$ ($\sin2\beta \gtrsim 0.7$).
The second population, Region~II, has $\lambda \lesssim 0.08$,
$\kappa/\lambda$ well below unity, and larger values of $\tan\beta \gtrsim 5$
($\sin2\beta \lesssim 0.38$).  For values of $\lambda$ between
the large and small values taken on within the disjoint green regions,
we find that the SM-like $h^0$ Higgs boson is too light to satisfy
the LEP bounds.  In the larger $\lambda$ portion, Region~I, 
the singlet $F$-term becomes important for smaller 
$\tan\beta$ and is responsible for increasing the mass of the $h^0$.  
At smaller $\lambda$, in Region~II, $m_{h^0}$ is very close to 
the value it would have in the MSSM, and larger values of 
$\tan\beta$ are necessary to push it above the LEP~II bound.  
We discuss the phenomenology of Regions~I and II below.

\subsection{Region~I: Higgs Decays to Pseudoscalars}

  Region~I consists of points with $\lambda \sim \kappa \gtrsim 0.4$
and $\tan\beta \lesssim 2.5$.  The Higgs phenomenology in this region
can be very different from the SM when 
$m_{a_s} < m_{h^0}/2$ and the $h^0$ state decays predominantly 
into pairs of $a_s$ pseudoscalars.  Evidently this requires a light
$a_s$ pseudoscalar and a sizeable effective coupling $c$.  
When this is not the case, the Higgs phenomenology
turns out to be very similar to the MSSM at large $m_{A^0}$.

\begin{figure}[ttt]
\vspace{1cm}
\begin{center}
        \includegraphics[width = 0.5\textwidth]{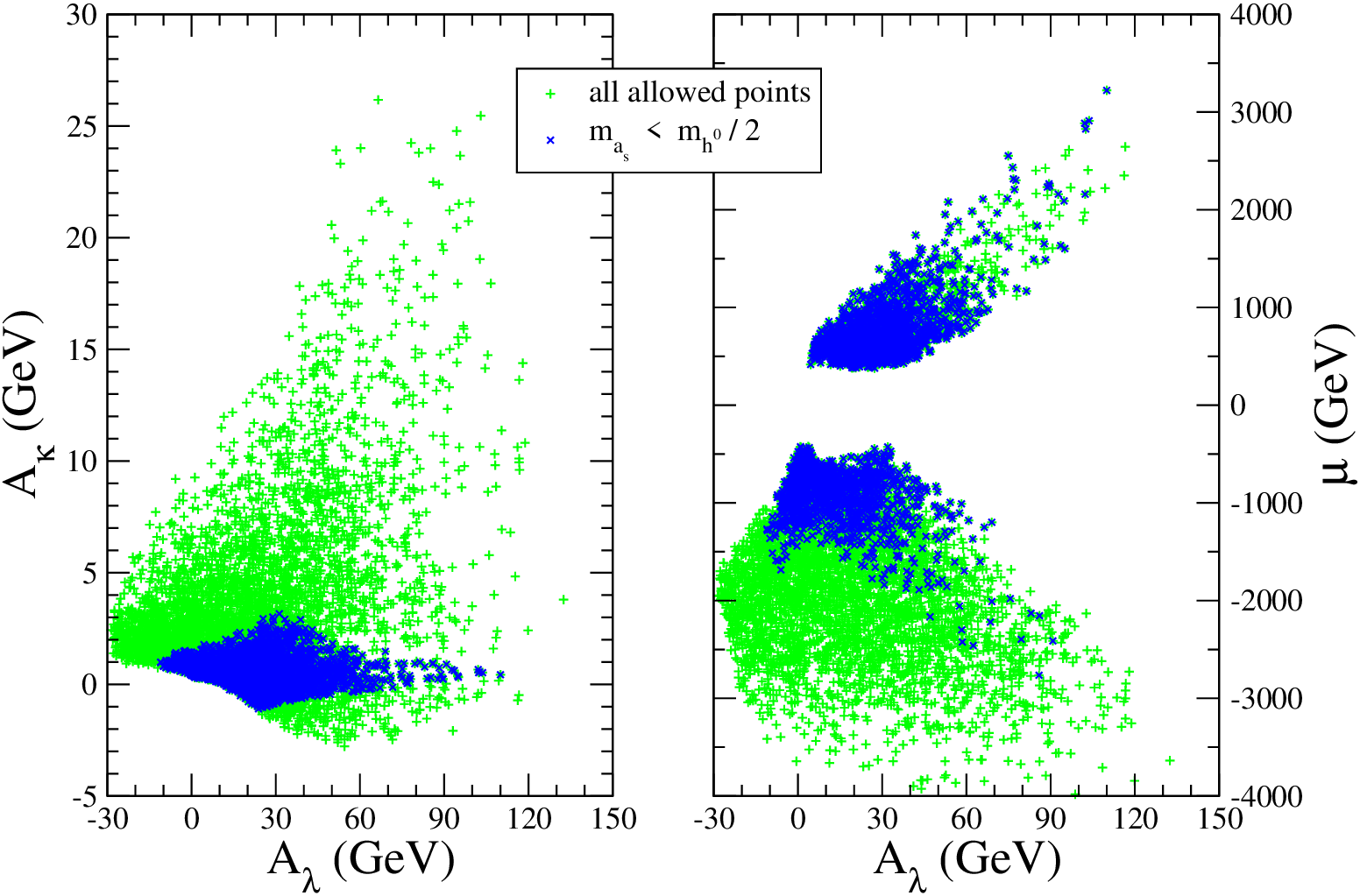}
\end{center}
\caption{Values of $A_{\lambda}$, $A_{\kappa}$, and $\mu$ among 
phenomenologically acceptable parameter points in Region~I described 
in the text.  The green (light gray) points satisfy all collider 
phenomenological constraints, while the blue (dark gray) points 
also have a light pseudoscalar with $m_{a_s} < m_{h^0}/2$.
}
\label{gauge-ms-al1}
\end{figure}

  As discussed in Section~\ref{u1r}, a light $a_s$ pseudoscalar 
can emerge from the spontaneous breakdown of an approximate $U(1)_R$ 
symmetry in the singlet sector.  This $U(1)_R$ is broken explicitly
by the singlet trilinear $A$ terms, but remains a good approximate
symmetry provided they are much smaller than $\mu$.  In Fig.~\ref{gauge-ms-al1}
we plot the values of $A_{\lambda}$ against $A_{\kappa}$, as well as 
$A_{\lambda}$ against $\mu$ for points within Region~I.
The green (light) points are consistent with all phenomenological bounds,
while the blue (dark) points also have $m_{a_s} < m_{h^0}/2$.
From the right panel of this figure we see clearly a hierarchy 
between the singlet $A$ terms and $\mu$.

  It is not hard to understand how the approximate $U(1)_R$ can 
give rise to $m_{a_s} < m_{h^0}/2$ by comparing Fig.~\ref{gauge-ms-al1}
with the results of Eq.~\eqref{cpodd1} and Eq.~\eqref{cpeven1}.\footnote{
Numerically, we find the tree-level expression of Eq.~\eqref{cpodd1} 
to be accurate only for larger values of $m_{a_s}$ due to additional 
quantum corrections.  However, Eq.~\eqref{cpodd1} remains useful for 
determining when the $a_s$ state will be light.}
To obtain such a light pseudoscalar, very small values 
of $A_{\kappa}$ are necessary.  There is a contribution to $m_{a_s}$ 
that goes as $\mu\,A_{\kappa}$, whereas the mass of the light 
$h^0$ Higgs boson is largely controlled by $\lambda v$ in this region.  
Recall that $\lambda^2v^2/\mu^2 \ll 1$. 
It is also necessary to have $A_{\lambda}/\mu$ reasonably small, 
although this requirement is much less severe due to the additional 
suppression by $\lambda^2v^2/\mu^2 \ll 1$ in Eq.~\eqref{cpodd1}.

  Among the parameter points with a sufficiently light $a_s$, 
the dominant contribution to the coupling $c$ in Eq.~\eqref{cform}
comes from the term involving $A_{\lambda}$.  
Thus, the small amount of $U(1)_R$ breaking induced by the 
gaugino masses and transmitted to $A_{\lambda}$ in the course of RG running
plays a dual role.  It must be small enough to keep the 
pseudoscalars light, but still large enough to facilitate 
$h^0\to a_sa_s$ decays.  As shown here, and previously 
observed in \cite{Dermisek:2006wr} (there with breaking at the GUT scale), 
the size of the $A$ terms derived from running can provide the 
right amount of $U(1)_R$ breaking to accomplish both of these tasks.

\begin{figure}[ttt]
\vspace{1cm}
\begin{center}
        \includegraphics[width = 0.5\textwidth]{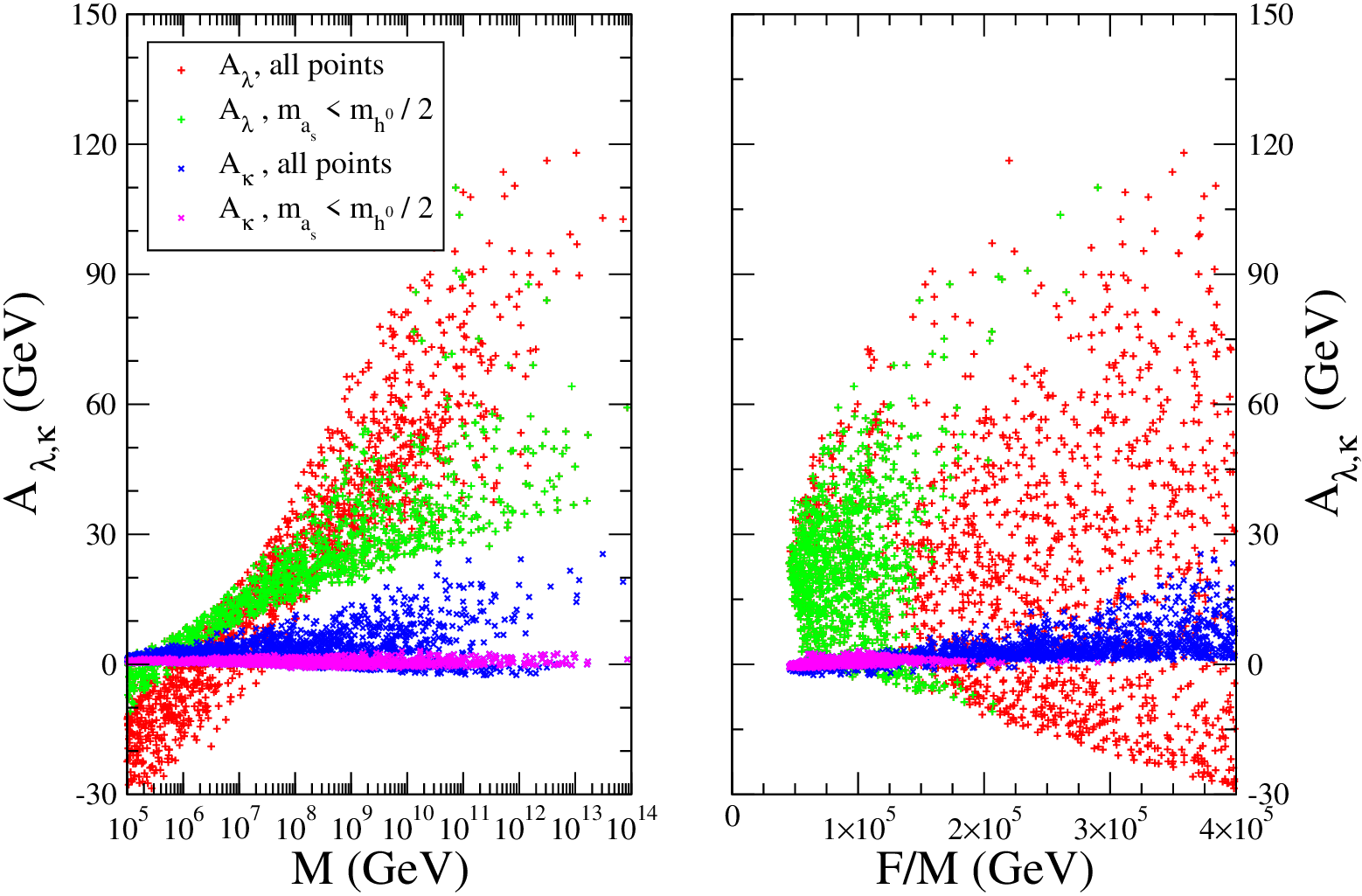}
\end{center}
\caption{Dependence of the singlet trilinear $A_{\lambda}$ and $A_{\kappa}$
terms on the messenger scale $M$ and the scale of supersymmetry breaking
$F/M$.  We exhibit phenomenologically allowed parameter points,
as well as points satisfying the additional condition of $m_{a_s} < m_{h^0}/2$.
}
\label{gauge-ms-fm}
\end{figure}

  The low-scale values of $A_{\lambda}$ and $A_{\kappa}$ are generated
in the course of RG running down from the messenger scale.
They are sourced indirectly by the gaugino masses.  Thus, the precise
values of $A_{\lambda}$ and $A_{\kappa}$ near the electroweak scale
are sensitive to the GMSB parameters $F/M$ and $M$.  
In Fig.~\ref{gauge-ms-fm} we show the values of $A_{\lambda}$ and $A_{\kappa}$
among allowed parameter points in Region~I as functions of both $F/M$
and $M$.  We exhibit allowed points both with and without 
$m_{a_s} < m_{h^0}/2$.  Not surprisingly, smaller values of the trilinear 
$A$ terms are obtained from lower values of $F/M$, corresponding 
to a lighter superpartner
spectrum.  For higher messenger scales $M$, the logarithmic enhancement 
of the singlet $A$ terms becomes stronger, leading to larger values 
of these couplings.  This tends to push up the mass of the $a_s$
pseudoscalar, but it also helps to enhance the decay width
of $h^0\to a_sa_s$ allowing this mode to dominate over decays to bottom quarks.

\begin{figure}[ttt]
\vspace{1cm}
\begin{center}
        \includegraphics[width = 0.5\textwidth]{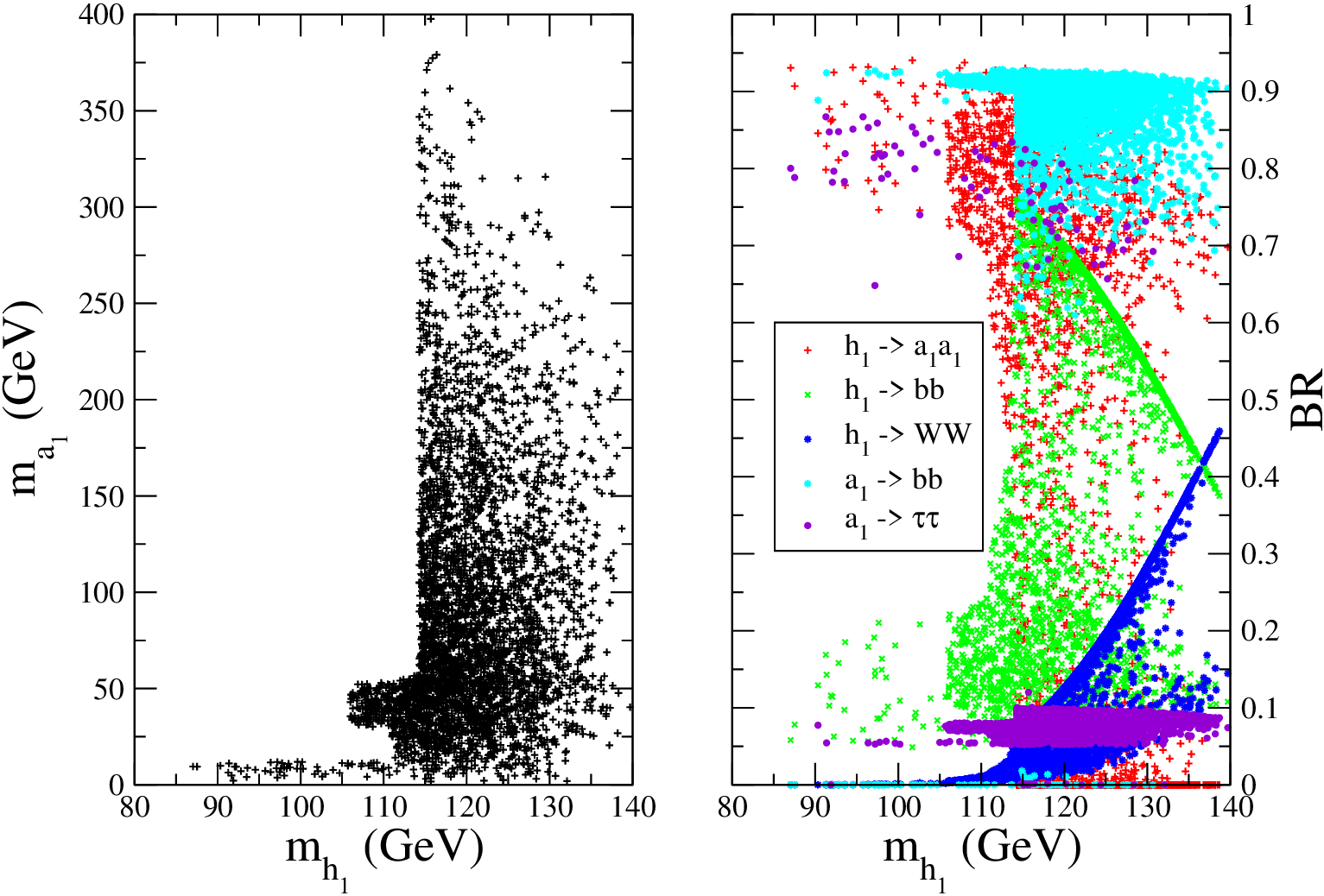}
\end{center}
\caption{Masses and branching fractions of the lighter $h^0$
and $a_s$ Higgs bosons for allowed parameter points in Region~I
described in the text.
}
\label{gauge-ms-mh1}
\end{figure}

In Fig.~\ref{gauge-ms-mh1} we show the masses of the lightest scalar $h^0$ 
and the lightest pseudoscalar $a_s$ for allowed parameter 
points in Region~I in the left panel.  In the right panel we show 
the dominant branching fractions for these states. 
We see that the decay properties of the $h^0$ Higgs differ 
markedly from those of a SM Higgs when the branching fraction
for $h^0\to a_sa_s$ is close to unity.  When this is the case,
the $h^0$ state can be significantly lighter than $114\,\gev$ and
still be consistent with the bounds from LEP~II, as discussed in
Ref.~\cite{Dermisek:2005ar}.  For pseudoscalar masses larger than about
$10\,\gev$, the $a_s$ decays primarily into $b\bar{b}$,
weakening the bound on the $h^0$ mass  to roughly $110\, \gev$.
Very light pseudoscalars, below about $10\,\gev$ in mass, tend
to decay mostly into $\tau\bar{\tau}$ (unless they are extremely light),
allowing for $h^0$ masses as low as about $90\,\gev$ to be consistent
with the bounds from LEP~II~\cite{Schael:2006cr}.
In the present context, it is challenging to make $A_{\kappa}$ small
enough to get such a light $a_s$.  We find that $m_{a_s} < 10\,\gev$ 
generally requires a cancellation between the $A_{\kappa}$ and  
$A_{\lambda}$ terms in Eq.~\eqref{cpodd1}, implying
a degree of fine-tuning.  
This possibility is constrained by and will be further probed 
by searches for $\Upsilon\to a_s\gamma$~\cite{Dermisek:2006py,cleonew}
as well as measurements of $(g-2)_{\mu}$~\cite{Domingo:2008bb}.

  While the decays of the SM-like $h^0$ Higgs boson can be significantly
modified by the presence of a light singlet pseudoscalar, 
the other properties of the $h^0$ state such as its production rates and 
decay widths into SM final states generally remain nearly unchanged 
relative to the SM Higgs.  For example, the mixing between 
the SM-like combination of the $CP$-even gauge eigenstates 
(\textit{i.e.} the combination that has the same couplings 
with the gauge bosons as in the SM) 
with the singlet in the $h^0$ is given by the matrix element
\beq
U_{12} \simeq -\frac{\lambda}{2\kappa}
\frac{\lambda\,v}{\mu}\left(\frac{\lambda}{\kappa}-\stb\right).
\eeq
Here we have kept only the leading terms in the expansion 
in $\lambda\,v/\mu$ and $A_{\lambda,\kappa}/\mu$.  
More details about the mixing matrix are given in Appendix~\ref{appa}.
Numerically, we find $|U_{12}| \lesssim 0.10$,
along with $|U_{11}| \gtrsim 0.995$.  In particular,
the dominant Higgs production channels 
via loop-induced coupling to gluons or direct couplings to gauge bosons
will only be reduced by a factor of $U_{11}^2$, or less than about $1\%$.

  The LHC signatures of the $h^0$ Higgs depend strongly on 
its branching fraction into pairs of $a_s$.
If these decays are suppressed, either by kinematics or a small
coupling, the Higgs signatures will be very similar to the MSSM.
When $BR(h^0\to a_sa_s)$ is non-trivial, the branching fractions 
of the $h^0$ into MSSM final states will be reduced according to
\beq
BR_i = BR_i^{MSSM}\,\left[1-BR(h^0\to a_sa_s)\right].
\eeq
Despite this suppression, some of the more promising SM Higgs
channels such as $h^0\to \gamma\gamma$ ($m_{h^o} \lesssim 125\,\gev$)
and $h^0\to WW^*$ ($m_{h^0} \gtrsim 125\,\gev$) could still be
visible in their own right even with $BR(h^0\to a_sa_s)$ close
to its maximal value of about $0.9$.  

  It may also be possible to search for the $h^0$ Higgs boson
through its decays to light pseudoscalars if both $BR(h^0\to a_sa_s)$
and $BR(a_s\to b\bar{b})$ are close 
to unity~\cite{Cheung:2007sva,Carena:2007jk}. 
This search relies on Higgs production in association with a weak 
gauge boson to reduce the background.  It  
is challenging because it requires multiple $b$-tags
to reduce the background, and therefore requires a high $b$-tag
efficiency along with a good understanding of the mis-tagging rate.
Even so, Ref.~\cite{Carena:2007jk} finds that it should be possible
to discover the $h^0$ Higgs through this channel with at
least 30 fb$^{-1}$ of integrated luminosity.  When the $a_s$
pseudoscalar is so light that it decays predominantly into $\tau\bar{\tau}$,
which we find to be fairly unlikely in Region~I,
 Ref.~\cite{Forshaw:2007ra} proposes a LHC search strategy using
forward proton tagging (central exclusive production) with
the proposed FP420 detectors~\cite{Albrow:2008pn}.
The recent study of Ref.~\cite{Belyaev:2008gj} suggests
that this decay channel might also be observable at the
LHC in the weak boson fusion production channel.

  We have concentrated so far on the phenomenology of the $h^0$
Higgs boson and the $a_s$ pseudoscalar.  The other Higgs states, 
the $CP$-even $H^0$ and $h_s$ along with the $CP$-odd $A^0$ and
the charged $H^{\pm}$, could also turn up at the LHC.
However, among the phenomenologically allowed points
in Region~I these states are all heavier than about $500\,\gev$
and mix only very weakly with the SM-like $h^0$.   This makes them
difficult to detect at the LHC.
Since $\tan\beta$ is small in Region~I, 
these states have significant branching fractions for
decays into $t\bar{t}$.  The $h_s$ scalar can also decay efficiently
into pairs of $a_s$ pseudoscalars, while the $A^0$ pseudoscalar
can also decay appreciably to the lighter superpartners.  
Given the large masses of these states
along with their weak production cross-sections, we do not expect
that they will produce a significant signal at 
the LHC~\cite{atlastdr,Ball:2007zza}.
The charged $H^{+}$ Higgs decays primarily into $t\bar{b}$,
as well as $h^0W^+$, and may possibly be visible through 
the former mode~\cite{atlastdr,Ball:2007zza}.  

  Besides the new Higgs bosons, the singlet superfield 
$S$ also gives rise to an additional fifth neutralino state.  
For the parameter values in Region~I, this state 
consists primarily of the fermion component of $S$
with a small higgsino admixture, and has a mass 
close to $2\kappa\mu/\lambda$.  As $\kappa/\lambda \sim 1$
and $|\mu| > M_{1,2}$ in Region~I, this state is typically
the heaviest neutralino.  Moreover, its small mixing with
the MSSM neutralinos implies that it will have a low
production rate at the LHC, either by direct creation
or through SUSY cascade decays.  Cascade decays could also 
provide an interesting source of $h^0$ Higgs bosons,
although we do not pursue this possibility here.

\subsection{Region~II: (Partial) Singlet Decoupling}

  This region corresponds to the phenomenologically allowed
points in Fig.~\ref{gauge-ms-lk} with $\lambda \lesssim 0.1$, 
$\kappa/\lambda \lesssim 0.4$, and $\tan\beta \gtrsim 5$
($\sin2\beta \lesssim 0.4$). 
As in Region~I, the minimal GMSB boundary conditions 
lead to values of $|\mu|$ that are much larger than 
$\lambda\,v$ and the singlet trilinear $A_{\lambda,\kappa}$ couplings.  
This signals a certain amount of fine-tuning in the electroweak 
symmetry breaking conditions.  From the above values and the 
minimization conditions of Eqs.~(\ref{musoln},\ref{kapsoln},\ref{mssoln}), 
we see that this region has $|m_S^2/\mu^2| \ll 1$ near the electroweak scale.
The small values of $\lambda \ll 1$ that arise in Region~II
are needed to satisfy the LEP bound on the SM-like Higgs boson mass.
Expanding Eq.~\eqref{cpeven1} in small $\kappa/\lambda$ 
and $\sin2\beta$, we find that 
\beq
m_{h^0}^2 \simeq \lambda^2v^2\left(\frac{{g'}^2+{g}^2}{2\lambda^2}
- \frac{\lambda^2}{\kappa^2}\right).
\eeq
To obtain a sufficiently large Higgs mass, 
it is then necessary that $\bar{g}^2/2\lambda^2 \gg \lambda^2/\kappa^2$, 
forcing $\lambda^2 \ll 1$.  

\begin{figure}[ttt]
\vspace{1cm}
\begin{center}
        \includegraphics[width = 0.5\textwidth]{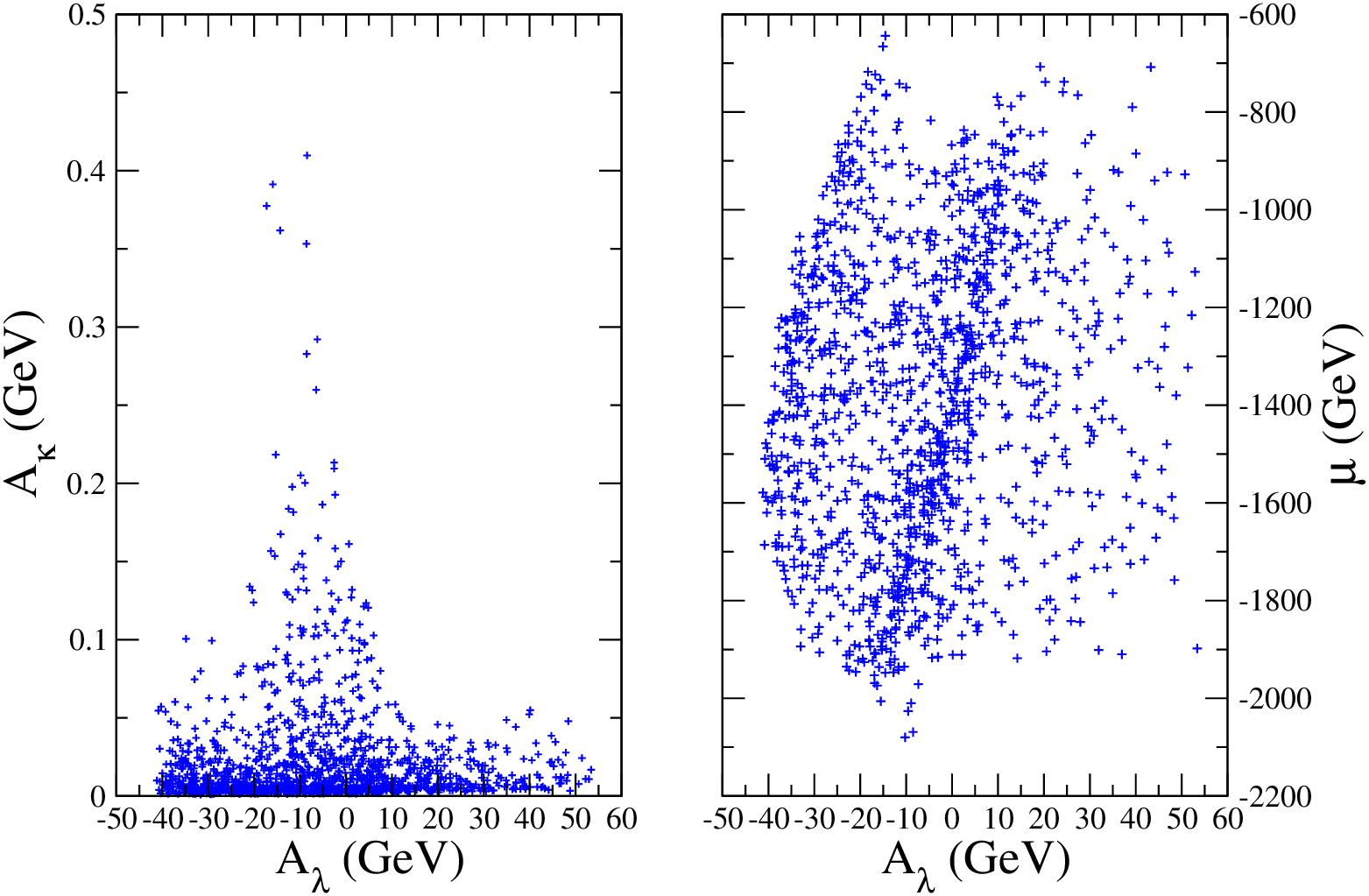}
\end{center}
\caption{Values of $A_{\lambda}$, $A_{\kappa}$, and $\mu$ among 
phenomenologically acceptable parameter points in Region~II described 
in the text.
}
\label{gauge-ms-al2}
\end{figure}

  The Higgs sector in Region~II always contains a very 
light pseudoscalar $a_s$ as well as a SM-like $h^0$ Higgs boson. 
The pseudoscalar mass is protected by an approximate $U(1)_R$ 
symmetry in the singlet sector due to the GMSB boundary conditions, 
in addition to an approximate $U(1)_{PQ}$ arising from the
small value of $\kappa/\lambda$.  On account of this double-protection,
the electroweak scale value of $A_{\kappa}$ is particularly small
among the allowed parameter points. This is clearly shown in the left 
panel of Fig.~\ref{gauge-ms-al2}.  From the right panel of 
Fig.~\ref{gauge-ms-al2} we also see that
the value of $|\mu|$ is much larger than $A_{\kappa}$, 
$A_{\lambda}$, and $\lambda\,v$ within the allowed parameter space.

\begin{figure}[ttt]
\vspace{1cm}
\begin{center}
        \includegraphics[width = 0.5\textwidth]{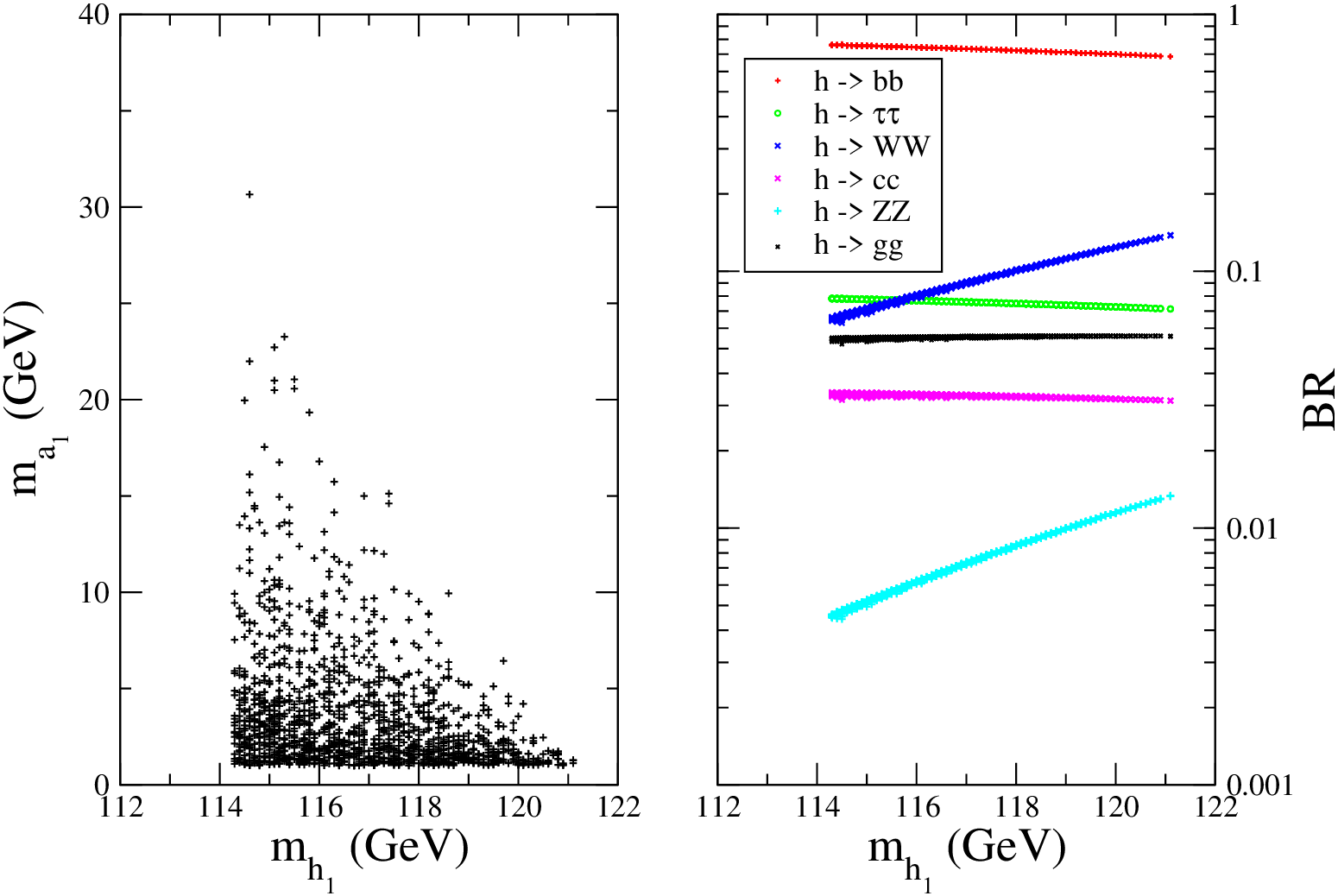}
\end{center}
\caption{Masses of the lighter $h^0$ and $a_s$ Higgs bosons,
and the dominant branching fractions of the $h^0$ Higgs 
for allowed parameter points in Region~II
described in the text.}
\label{gauge-ms-mh2}
\end{figure}

  With these very small values of $A_{\lambda}$ and $A_{\kappa}$, 
the mass of the pseudoscalar $a_s$ in Region~II is always much 
less than half the mass of the SM-like $h^0$ Higgs boson.  
This feature is illustrated in the left panel of Fig.~\ref{gauge-ms-mh2}.  
In the right panel of Fig.~\ref{gauge-ms-mh2} we show the dominant 
branching fractions of the $h^0$ Higgs boson.  
Despite being kinematically allowed, we find that $h^0\to a_sa_s$ almost 
never occurs, with $BR(h^0\to a_sa_s)<  3\times 10^{-4}$.  
Instead, the decay fractions of the $h^0$ Higgs are nearly identical 
to those of a SM Higgs. The reason for this suppression of $h^0\to a_sa_s$
can be seen in Eq.~\eqref{cform}; the coupling $c$ is proportional 
to $\lambda^2 \ll 1$.

  Collider production of the $h^0$ Higgs in Region~II is also nearly
identical to the SM.  The singlet component of this mass eigenstate
is quite small, corresponding to the mixing element $|U_{12}| \lesssim 0.25$,
while the suppression of Higgs production through the
usual channels is down by only $U_{11}^2 \gtrsim 0.94$.
(See Appendix~\ref{appa} for a full account of the mixing matrices.)
Thus, the LHC signatures of the $h^0$ in Region~II will be nearly
identical to those of a SM Higgs (or an MSSM Higgs at large $M_{A^0}$).
The mixing among the $CP$-odd states is even more suppressed,
with the light $a_s$ being almost exclusively singlet.
This provides another way to understand why the branching fraction
of the $h^0$ into pseudoscalar pairs is so small in Region~II.

  Among the heavier Higgs bosons, the mostly-singlet 
$h_s$ can be relatively light since its mass is suppressed 
by a factor of $\kappa/\lambda \ll 1$.  We find masses as 
low as $170\,\gev$, and as high as $1500\,\gev$.  While this
state is mostly singlet, a mixing with the SM-like Higgs as
large as $|U_{21}| \simeq 0.25$ is possible.  The primary
decay modes are $WW$, $ZZ$, and $h^0h^0$.  For larger mixings and 
moderate masses, in the range $200\,\gev \lesssim m_{h_s} 
\lesssim 500\,\gev$, the $h_s$ may be visible at the LHC 
through its $ZZ$ final state. 
All the other Higgs bosons, the $H^0$, $A^0$, and $H^{\pm}$,
have masses in excess of $800\,\gev$, and are challenging 
to find at the LHC.

  The rest of the particle spectrum within Region~II is similar
to the MSSM, but with an additional neutralino from the
fermion component of $S$.  To push $m_{h^0}$ above
the LEP-II bound, the superpartner spectrum must be somewhat
heavy with $M_3 \gtrsim 800\,\gev$.  
On the other hand, the corresponding mostly-singlet neutralino state
state can be as light as $170\,\gev$ in Region~II,
and can even be the lightest superpartner aside from the gravitino.  
Due to the very small values of $\lambda$ and $\kappa$, 
this state is almost pure singlet with a tiny higgsino component, 
and couples only very weakly to the rest of the MSSM.  It will
therefore almost completely decouple from the LHC phenomenology
unless it is the NLSP (with a gravitino LSP).  In this case,
the presence of an additional state can modify cascade decay
chains~\cite{Ellwanger:1997jj,Kraml:2005nx,Barger:2006kt} 
and possibly also give rise to neutral displaced 
vertices from decays of the NNLSP to 
the NLSP~\cite{Ellwanger:1997jj,Hesselbach:2000qw}.
A mostly singlet NLSP could also be problematic
if it decays after the onset of nucleosynthesis~\cite{Feng:2003uy,
Ellis:2003dn,Kawasaki:2004qu}. 
While a full analysis of this issue is beyond the scope of 
the present work, we expect that lighter gravitino masses 
(leading to shorter decay times) might be necessary to ensure
that these decays occur before the light elements are formed.

\section{Gauge Mediation in the NMSSM with Exotics \label{gauge}}

As a second variation on minimal gauge mediation in the NMSSM, we consider 
adding charged vector-like exotics to the 
theory~\cite{Dine:1993yw,deGouvea:1997cx,Agashe:1997kn}.  
We study exotics in the form of $\tilde{D}\oplus \tilde{D}^c$,
and $\tilde{L}\oplus \tilde{L}^c$, with $\tilde{D} = (3,1,-1/3)$ 
and $\tilde{L} = (1,2,-1/2)$.  Taken together, these exotics have 
the quantum numbers of a $\ffb$ of $SU(5)$.  We include the 
trilinear superpotential couplings 
\beq
W \supset \xi_{D}\,S\,\tilde{D}\,\tilde{D}^c 
+ \xi_L\,S\,\tilde{L}\,\tilde{L}^c,
\label{newyuk}
\eeq
as well as a corresponding set of soft supersymmetry breaking operators.
These couplings tend to drive $m_S^2$ negative in the course of RG running
from the gauge messenger scale down to near the electroweak scale, 
thereby facilitating singlet condensation and electroweak symmetry breaking.
The relevant RG equations for this evolution with $N_D$ sets of 
$\tilde{D}\oplus\tilde{D}^c$ and $N_L$ sets of $\tilde{L}\oplus\tilde{L}^c$
are listed in Appendix~\ref{appb}.  Note that we assume the exotics 
do not act as gauge messengers, which can be enforced with the 
approximate $\mathbb{Z}_3$ symmetry discussed in the introduction.

\subsection{Allowed Parameter Regions}

  We have searched for regions of the NMSSM parameter space that
lead to an acceptable phenomenology by scanning over the model parameters.  
Our strategy consists of specifying $\lambda$, $\xi_D$, $\xi_L$, 
and $F/M$ at the GMSB messenger scale $M$, 
along with $\tan\beta$ near the electroweak scale,
and computing the low-energy spectrum that results from these input
parameters.  In doing so, we use a modified version of 
NMSPEC/NMHDECAY~\cite{Ellwanger:2004xm}
to perform the RG evolution of the model parameters, to find the low-scale 
values of $\kappa$, $|\mu|$, and $m_S^2$, as in Eqns.~(\ref{musoln},
\ref{kapsoln},\ref{mssoln}), and to determine the phenomenological
constraints.  The value of $m_S^2$ computed
in this way will not usually agree with the GMSB boundary condition of
$m_S^2(M) \simeq 0$.  To correct for this, we adjust the value of 
$\xi_D$ (or $\xi_L$), repeat the running, and iterate until the
value of $|m_S^2|$ at the input scale $M$ lies below a small cutoff value.
The parameter ranges covered in our scans are
\beq
\begin{array}{ccc}
\lambda\in [0.001,0.7],&\xi_L\in [0,1],&\xi_D \in [0,1],\\
&&\phantom{.}\\
\tan\beta\in [1,50],&M \in [10^5,10^{14}]\gev,&
~~F/M \in [2,40]\times 10^4\,\gev.
\end{array}
\eeq
We assume minimal GMSB boundary conditions for all the 
soft terms with a single set of $\ffb$ messengers.

\begin{figure}[ttt]
\vspace{1cm}
\begin{center}
        \includegraphics[width = 0.5\textwidth]{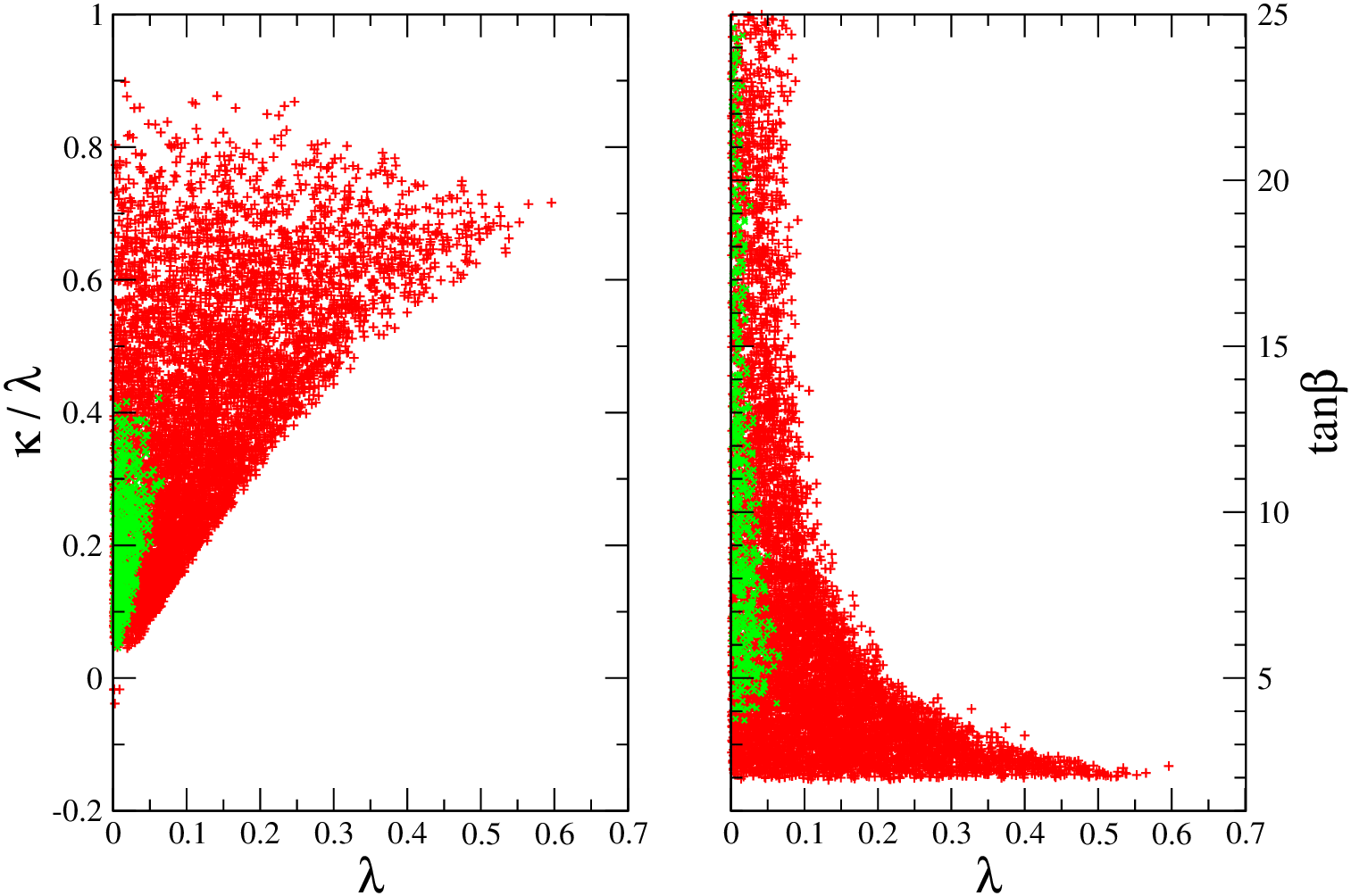}
\end{center}
\caption{Allowed parameter points for the NMSSM with 
minimal GMSB boundary conditions for a single set of $\ffb$ messengers,
and a single set of $\ffb$ exotics.  The red (dark) points do not include 
Higgs constraints while the green (light) points do.
}
\label{gauge-ex-lk}
\end{figure}

  In Fig.~\ref{gauge-ex-lk} we show the allowed parameter points 
obtained by scanning with a single set $N_5 = 1$ 
of $\ffb$ (non-messenger) exotics.
In the left panel of this figure we show points in the 
$\lambda$-$\kappa$ plane, while in the right panel we exhibit
points in the $\lambda$-$\tan\beta$ plane.  The red (dark) points agree 
with all relevant phenomenological bounds except for the mass 
constraints on the neutral $CP$-even and $CP$-odd Higgs bosons, 
while the green (light) points are also consistent with the 
Higgs constraints.  As in the previous section,
we do not demand the lightest neutralino be the LSP: in 
gauge mediation the gravitino is usually 
expected to be the true LSP. We do require that the couplings 
remain perturbatively small up to the scale $M$ (but not $M_{GUT}$).

  The phenomenologically acceptable parameter points 
in Fig.~\ref{gauge-ex-lk} 
are similar to the parameter points found in Region~II discussed in 
the previous section.  These points all have very small values of 
$\lambda$, and relatively small values of $\kappa/\lambda$ and $\sin2\beta$.  
Small values of $\lambda$ in this region are needed
to make the SM-like Higgs boson sufficiently heavy.
Unlike in the previous section, however, we do not find any
points at larger values of $\lambda$ and $\kappa$, analogous to Region~I .  
For points of this type to lead to an acceptable pattern of symmetry breaking,
Eqs.~(\ref{musoln},\ref{kapsoln},\ref{mssoln}) require 
$|m_S^2|/\mu^2 \sim 1$ (or large $A$ terms).  
While the new Yukawa couplings in Eq.~\eqref{newyuk} help to drive
$m_S^2$ to more negative values, their effect is not strong enough
to open a region of parameter space with $\lambda \sim \kappa \sim 1$.
With a single set of non-universal exotics,
such as a lone $\tilde{L}\oplus \tilde{L}^c$ 
or $\tilde{D}\oplus \tilde{D}^c$, we find 
qualitatively similar results.  

  To magnify the effect of the exotics on the running of $m_S^2$,
we have also performed scans with multiple sets of $\ffb$ 
(non-messenger) exotics.
For simplicity, we assume universal values of the couplings
$\xi_D$ and $\xi_L$ for all flavors of exotics.  
Adding even only a second set of $\ffb$ exotics 
opens a new region of phenomenologically consistent 
points, similar to Region~I discussed in Section~\ref{free}.  
These points have $\kappa \sim \lambda \gtrsim 0.4$
and $\tan\beta \lesssim 2.5$.  A similar region can 
emerge with more than two sets of $\ffb$ exotics.  
The allowed region with $\lambda \ll 1$ also remains.
With several sets of exotics, a tension arises between
larger values of $\lambda$ and $\kappa$ at the low scale
and perturbativity up to the messenger scale $M$ since the exotic 
Yukawa couplings of Eq.~\eqref{newyuk} speed up to the running of $\lambda$.  
 
\begin{figure}[ttt]
\vspace{1cm}
\begin{center}
        \includegraphics[width = 0.5\textwidth]{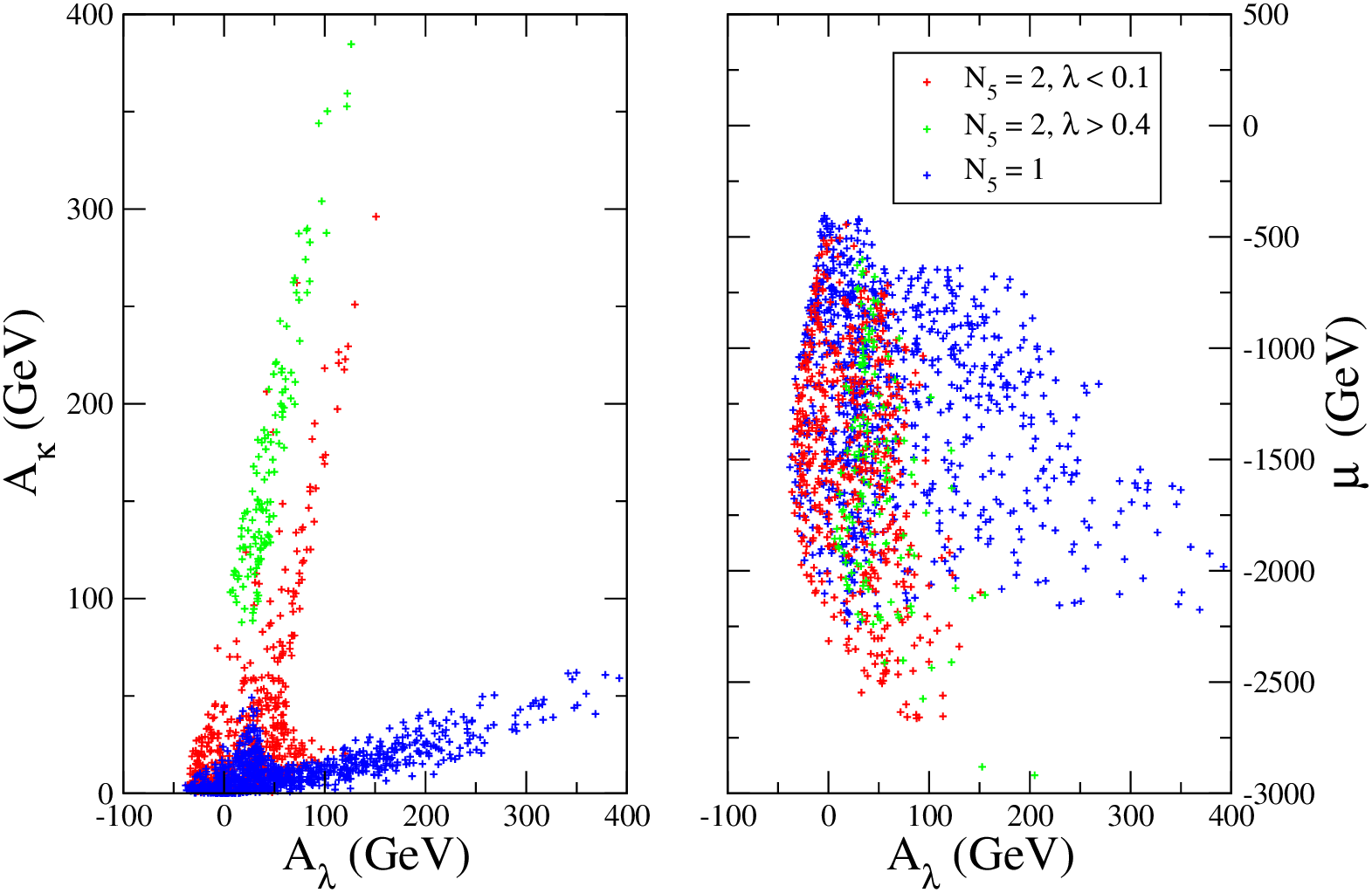}
\end{center}
\caption{Values of $A_{\lambda}$, $A_{\kappa}$, and $\mu$ among 
phenomenologically acceptable parameter points.  We exhibit 
allowed points with $N_5 =1$ and $\lambda < 0.1$, as well as
$N_5 = 1$ and $\lambda < 0.1$, and $N_5 = 2$ and $\lambda > 0.4$.
}
\label{gauge-ex-al2}
\end{figure}

  The NMSSM Higgs boson spectrum depends sensitively on
the values of $A_{\lambda}$, $A_{\kappa}$, and $\mu$ near
the electroweak scale.  In Fig.~\ref{gauge-ex-al2} we plot
these parameters for $N_5 = 1$ and $N_5 = 2$
sets of $\ffb$ exotics.  For the case of $N_5 = 2$, 
we split up the points according to whether they fall into
the $\lambda < 0.1$ region, or the $\lambda > 0.4$ region.
These plots indicate that $A_{\lambda}$ and $A_{\kappa}$
remain smaller than $|\mu|$ (larger values of $A_{\lambda,\kappa}$
correspond to very large values of $\mu$), but are enhanced relative
to the electroweak scale.  This enhancement comes from the contributions
from the new Yukawa couplings $\xi_D$ and $\xi_L$ in Eq.~\eqref{newyuk}
to $A_{\lambda}$ and $A_{\kappa}$ in the course of RG running.

\subsection{LHC Higgs Signatures with Exotics}

  Much like in Section~\ref{free}, the Higgs phenomenology with
additional vector-like exotics can be split into two classes,
depending on whether $\lambda$ is small ($\lambda \lesssim 0.1$)
or $\lambda$ is larger ($\lambda \gtrsim 0.4$).  This second
possibility only occurs with at least two sets of $\ffb$ exotics.
We will discuss both of these cases in turn.  

\begin{figure}[ttt]
\vspace{1cm}
\begin{center}
        \includegraphics[width = 0.5\textwidth]{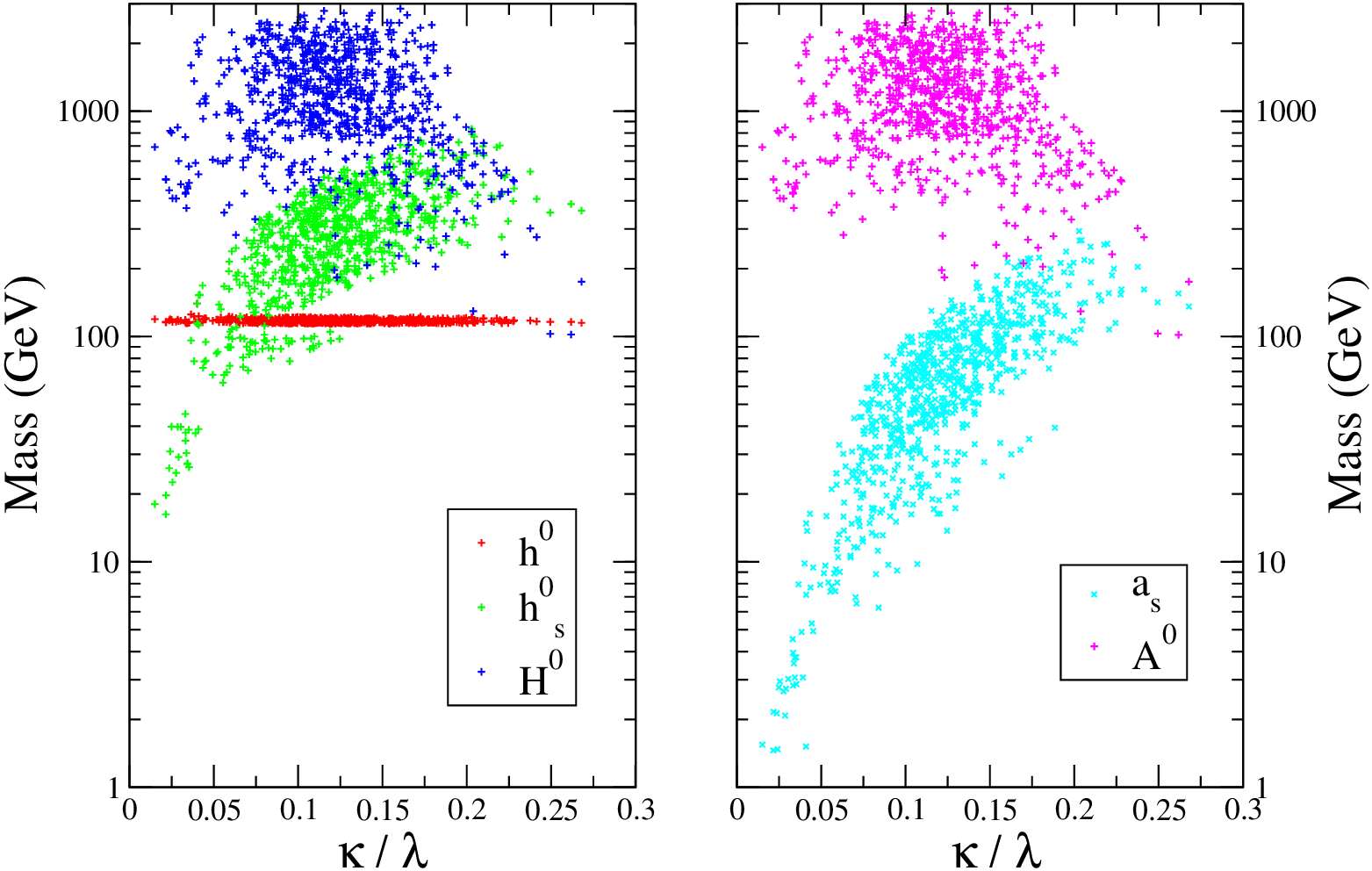}
\end{center}
\caption{Neutral Higgs boson masses for phenomenologically
allowed points in the NMSSM with a minimal GMSB spectrum
and a single set of $\ffb$ exotics.
}
\label{gauge-ex-mh}
\end{figure}

  In the allowed parameter region with $\lambda \ll 1$
(along with $|\kappa/\lambda| \ll 1$ and $\sin2\beta \ll 1$)
the Higgs boson phenomenology is quite similar to what we 
found in Region~II discussed in Section~\ref{free}.
We show the masses of the neutral $CP$-even and $CP$-odd Higgs
bosons in Fig.~\ref{gauge-ex-mh} for phenomenologically consistent
points obtained in our scan with a single set of $\ffb$ exotics.
The lighter Higgses consist of the SM-like $h^0$,
the mostly-singlet $h_s$, and a light mostly-singlet pseudoscalar $a_s$.
The $A^0$, $H^0$, and $H^{\pm}$ states are similar to their
MSSM counterparts, and are generally very heavy.
There are some slight but important differences compared to Region~II 
of Section ~\ref{free}, however.

  The singlet $A$ terms that
arise in the course of RG running are now somewhat larger, due to the 
new exotic Yukawa couplings (see Fig.~\ref{gauge-ex-al2}). The result is 
the mostly singlet pseudoscalar $a_s$ can be considerably 
heavier than in Region~II discussed above.
Depending on the values of these trilinear couplings,
$m_{a_s}$ can range from below $1\,\gev$ all the way up
to over $250\,\gev$.  This state is nearly pure singlet
with only a tiny admixture of the $A^0$ as a result of 
the very small values of $\lambda$ and $\kappa$.
We expect it to be almost completely decoupled from the
rest of the theory, and consequently invisible at the LHC. 
Similar considerations minimize the ability to probe this 
state through upsilon decays, if kinematically accessible.  
The dominant decay modes of the $a_s$ are into $b\bar{b}$, $\tau\bar{\tau}$, 
and so on, depending on which channels are kinematically accessible.  

  The LHC phenomenology of the $h^0$ and $h_s$ states is more interesting.
As in Region~II of the previous section, the branching fractions 
of the $h^0$ Higgs are similar to those of a SM Higgs, 
and thus the LEP bound of about $114\,\gev$ applies throughout 
most of the parameter region.  
An exception occurs when the mixing between the 
$h^0$ and $h_s$ states is enhanced when they become nearly 
degenerate.  Even in this case, the reduction in the mass bound 
on the SM-like state is usually less than a few $\gev$.  
The mostly-singlet $h_s$ Higgs can
be very light in some cases, with masses as low as $20\,\gev$,
but also as heavy as about $800\,\gev$.  This state inherits its 
couplings to the MSSM through its mixing with the SM-like Higgs.
Thus, the $h_s$ decay modes mirror those 
of a SM Higgs boson.  The chief exception to this occurs when 
the channel $h_s \to h^0h^0$ opens up.  The corresponding 
branching fraction can be as large as $BR(h_s\to h^0h^0) \simeq 0.4$.  
For heavier $h_s$ masses, this state might be visible at the 
LHC through its decays to $ZZ$, even though its production 
cross-section is suppressed relative to a SM Higgs 
by its large singlet component.  All the other Higgs boson states
are heavier than about $500\,\gev$, making them challenging
to find at the LHC unless $\tan\beta$ is very large.

  The rest of the particle spectrum within this small $\lambda$ region
is similar to that of the MSSM.  Compared to Region~II,
the overall superpartner scale can be somewhat lower,
with gluino masses as small as $M_3 \simeq 600\,\gev$ now possible,
and the additional mostly-singlet neutralino can be very light.
The mass of this singlet neutralino is approximately $2\kappa\,\mu/\lambda$,
which can be as light as about $20\,\gev$ for particularly small values
of $\kappa/\lambda$, but also as large as several hundred $\gev$
in other portions of the allowed parameter space.  Mixing between
the singlet fermion and the other neutralinos is always very small.
The dominant contribution comes from the higgsinos with
a mixing angle $|\mathcal{O}_{15}| \lesssim 0.01$.  Thus, this fifth 
neutralino state is nearly invisible at colliders unless 
it is the NLSP (with a gravitino LSP).  As discussed in Section~\ref{free},
a mostly-singlet NLSP can modify sparticle cascade 
chains~\cite{Ellwanger:1997jj,Kraml:2005nx,Barger:2006kt}.
In this case, lighter gravitino masses may be required to
ensure that the NLSP decays safely before nucleosynthesis.

  The second allowed region of NMSSM parameter space with 
heavy exotics and minimal GMSB boundary conditions has $\lambda \gtrsim 0.4$ 
and smaller values of $\tan\beta \lesssim 2.5$.  These parameter ranges 
are similar to those encountered in Region~I of Section~\ref{free}.
Despite the similarity of these values, the Higgs boson phenomenology does 
not mirror that of Region~I.  In fact, it does not 
significantly differ from that of the MSSM.
To populate this large $\lambda$ region, 
at least two sets of $\ffb$ exotics are required in order to drive 
$m_S^2$ sufficiently negative over the course of the RG evolution.  
The exotic Yukawa couplings, $\xi_D$ and $\xi_L$, responsible for doing 
so also contribute to the low-scale values of $A_{\lambda}$ and 
$A_{\kappa}$.  This ruins the approximate $U(1)_R$ symmetry in the 
singlet sector.  As a result, the singlet pseudoscalar is no 
longer very light, with masses above $m_{a_s} \gtrsim 350\,\gev$.  
The SM-like $h^0$ state is therefore very SM-like, both in 
its production and decay modes, since it is no longer able to
decay into pairs of the $a_s$ pseudoscalar.  All the other Higgs boson 
states are heavier than about $500\,\gev$, and will be difficult to find 
at the LHC, particularly with the smaller values of $\tan\beta$ 
that occur in this region.  The additional neutralino state 
that arises from the fermion component of the singlet $S$
is also quite heavy, with a mass of $2\,\kappa\mu/\lambda \sim 2\mu$, 
and has only a small mixing with the higgsinos.  It too will be
essentially invisible at the LHC.

\subsection{Phenomenology of the Exotics}

  The vector-like exotic states we have added to the
theory to facilitate electroweak symmetry breaking can themselves
be a source of new signatures at the LHC. 
These charged exotics must certainly be heavy enough to have
avoided detection already.  Beyond this, the exotics are
problematic for cosmology if they are overly long-lived (or stable).
We briefly consider the additional bounds and potential signatures
that arise from the exotics.  These should inform any model 
building attempt.

\begin{figure}[ttt]
\vspace{1cm}
\begin{center}
        \includegraphics[width = 0.7\textwidth]{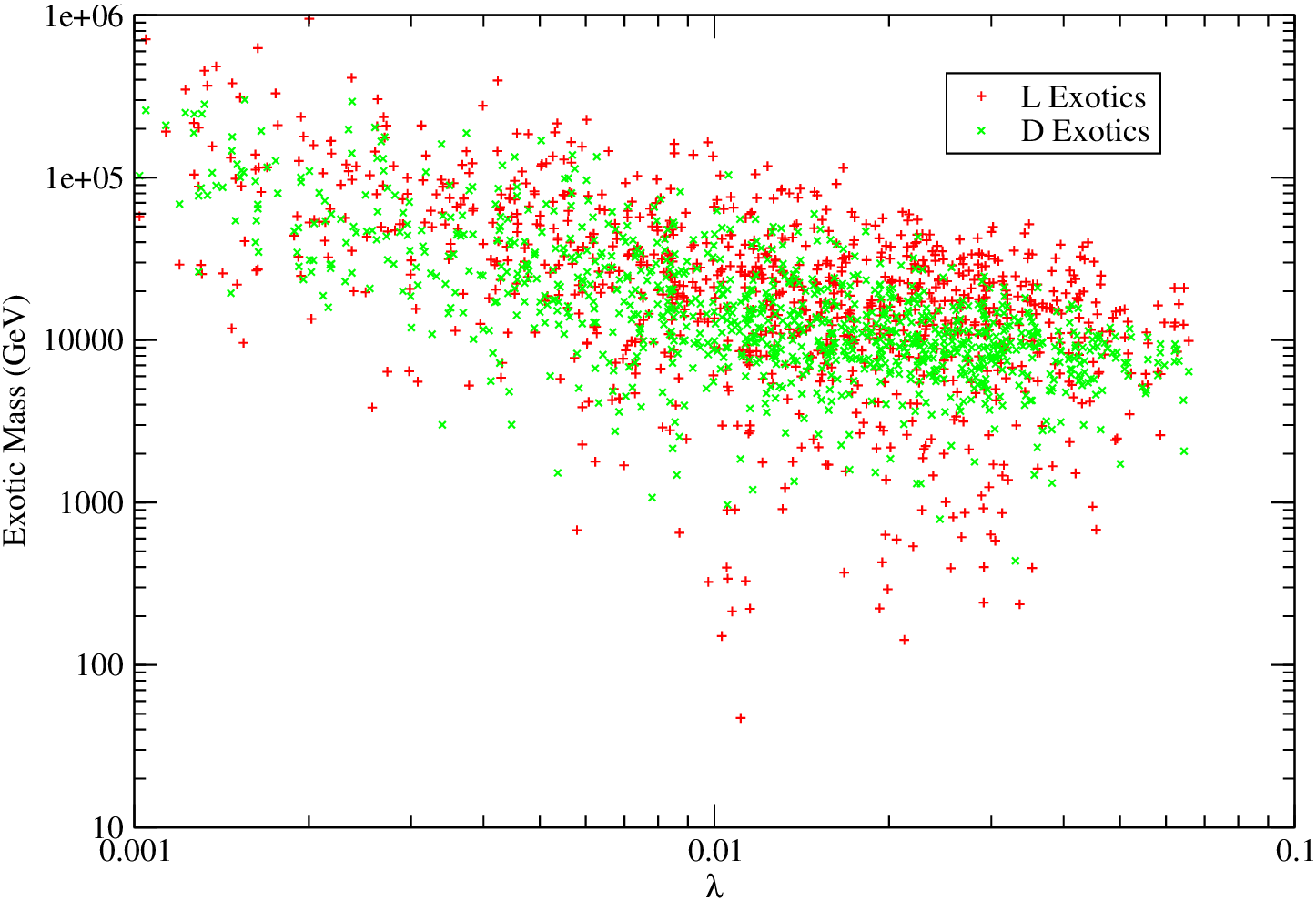}
\end{center}
\caption{Masses of the charged $D$ and $L$ exotics among the
phenomenologically allowed parameter points found for the NMSSM
with a single set of $\ffb$ exotics and a minimal GMSB soft mass spectrum.}
\label{mexotic}
\end{figure}

  In Fig.~\ref{mexotic} we plot the masses of the charged
$\tilde{D}$- and $\tilde{L}$-type exotics originating from 
the superpotential couplings of Eq.~\eqref{newyuk}.  As before, 
the red (dark) points in this plot are consistent with all 
phenomenological bounds other than the constraints on 
the Higgs sector (and on the exotics themselves), 
while the green (light) points satisfy all relevant Higgs bounds 
as well.  Nearly all the points in this figure have
exotic masses well in excess of the current limits.

  The precise collider bounds on the exotics depend on whether
they are long- or short-lived on collider timescales.
In the case of long-lived charged leptons, the best bound comes 
from searches by OPAL for (effectively) stable charged particles, 
and is $m_{\ell} \gtrsim 100\,\gev$ for a heavy spin-$1/2$ particle 
with electric charge $\pm1$ and no color~\cite{Abbiendi:2003yd}.
The bound for decaying charged leptons is about the same.
The most stringent bound on long-lived heavy quarks comes from
Tevatron searches for charged massive particles.
A preliminary analysis with $1\,fb^{-1}$ of Tevatron Run~II data suggests a
limit on the cross section for such states of about
$0.1$ pb ~\cite{tevachamp}, corresponding to quark masses
up to nearly $300\,\gev$. For short-lived heavy quarks, the precise
bound depends on how they decay. CDF has performed a preliminary search
for a heavy exotic top quark decaying through $t'\to Wq$, and find
$m_{t'} > 258\,\gev$~\cite{tprime}.  No $b$-tag is used in the analysis
so we expect this result to apply for a short-lived exotic 
$\tilde{D}$ as well~\cite{Kribs:2007nz}.

  To be cosmologically acceptable, the heavy exotics cannot be
too long-lived.  If stable on the lifetime of the universe, 
heavy charged exotics are very stringently constrained by
searches for anomalously heavy atoms.
These bounds are so severe that even the tiny density of heavy
exotics created by cosmic rays is unacceptably large~\cite{Byrne:2002ri}.
Long-lived heavy exotics are also dangerous if they decay after 
nucleosynthesis as their decay products can modify the light 
element abundances, distort the CMB blackbody spectrum, 
or contribute to cosmic rays~\cite{Kang:2007ib}.  On the other hand,
a significant coupling between the exotics and the MSSM matter
fields can give rise to too much flavor mixing.

  In the NMSSM with an approximate $\mathbb{Z}_3$ discrete symmetry, 
the exotic $\tilde{D}^{(c)}$ and $\tilde{L}^{(c)}$ states can
decay through the $d=4$ superpotential operators 
\beq
L\tilde{L}^c\,H_u\,H_d,~~L\tilde{L}^c\,SS,~~
D^c\tilde{D}\,H_u\,H_d,~~D^c\tilde{D}\,SS.
\label{exomix}
\eeq
Such operators can be consistent with the approximate
$\mathbb{Z}_3$ symmetry while still allowing all the standard NMSSM
operators, as well as the neutrino mass operator $(LH_u)^2$, and forbidding
mixing between the exotics and the MSSM matter states at the 
renormalizable level.  Decays through the operators
of Eq.~\eqref{exomix} occur safely before nucleosynthesis provided
the heavy mass scale suppressing them is less than about the
GUT scale, $M_{GUT} \simeq 10^{16}\,\gev$~\cite{Kang:2007ib,Arvanitaki:2005fa}.

  If the charged and colored exotics are not too heavy,
they might lead to observable signatures at the LHC.
Stable (on collider times) heavy quarks were studied 
recently in Ref.~\cite{Kang:2007ib}.  These can form charged
exotic hadrons that punch through to the muon chamber.  
By measuring the time of flight, they can be distinguished
from ordinary muons.  A significant signal of ten events with 
almost no background will be generated with $10\,fb^{-1}$ for 
heavy quark masses below $m_{D} \lesssim 1700\,\gev$, 
and $m_{\tilde{D}} \lesssim 1450\,\gev$ for the scalar superpartner.  
The precise limits on unstable heavy quarks depend on how they decay, 
but are generally of the similar size.  Heavy stable charged leptons 
can also be detected at the LHC through time-of-flight 
techniques~\cite{Allanach:2001sd,Fairbairn:2006gg}.  
Masses below about $950\,\gev$ can be detected in this way 
with 100 fb$^{-1}$ of integrated luminosity.

\section{Gaugino Mediation in the NMSSM\label{ino}}

  Gaugino mediation of supersymmetry breaking shares many of
the attractive features of gauge-mediated supersymmetry
breaking~\cite{Kaplan:1999ac,Chacko:1999mi}.  
In the context of obtaining a light pseudoscalar in the NMSSM, 
minimal gaugino mediation has the helpful property of
vanishing trilinear $A$ terms in the singlet sector at 
the mediation scale.  Unfortunately, like minimal gauge mediation, 
this mediation mechanism in its minimal form has trouble generating a
sufficiently negative low-scale value of the singlet soft mass $m_{S}^2$
to obtain an acceptable pattern of electroweak symmetry breaking.
In this section we study gaugino mediation in the NMSSM with 
the same modifications of the previous two sections:  we augment 
the theory by additional contributions to the input value of $m_S^2$, 
or by adding new vector-like exotics coupled to the $S$ field.

\subsection{Higgs Phenomenology with $m_S^2$ Free}

 We study first minimal gaugino mediation
within the NMSSM with the singlet soft mass $m_S^2$ taken
to be a free parameter.  One might motivate this scenario 
by putting the $S$ field in the bulk, so that it could feel 
the SUSY breaking directly.  However, to avoid generating singlet $A$
terms that would ruin the approximate $U(1)_R$ symmetry in the singlet
sector, such a bulk singlet should couple to a SUSY breaking source
that also preserves the $U(1)_R$.  It is a model-building challenge
to obtain such a source of supersymmetry breaking with the correct mass scale.

  As in Section~\ref{free}, we search for phenomenologically 
acceptable parameter regions by scanning over the model parameters.
Our scan encompasses the ranges
\bea
\lambda \in [0.001,0.7],&~~~\tan\beta \in [1,50]\,\gev,\\
M_{1/2}\in[100,1500]\,\gev,&~~M_c \in [10^{5},10^{17}]\,\gev.\nnmb
\eea
The parameter $M_{1/2}$ defines the gaugino mass at the 
compactification scale $M_c$ through the relation  
\beq
M_a(M_c) = 2\,g_a^2(M_c)\,M_{1/2}.
\eeq
This choice corresponds to universal gaugino masses
equal to $M_{1/2}$ when $M_c = M_{GUT}$ ($g_a^2(M_{GUT}) \simeq 1/2$).  
The input values of all the soft scalar masses (save $m_S^2$) and all the 
$A$ terms are taken to vanish at the input scale $M_c$.
Over the course of the RG running, we require that all couplings
remain perturbatively small up to $M_c$.

  The allowed regions found in our scans 
for minimal gaugino mediation with $m_S^2$ free are nearly identical to what 
we found in Section~\ref{free} for minimal GMSB, and illustrated in
Fig.~\ref{gauge-ex-lk}.  As before, there are two disjoint phenomenologically
consistent regions: one at small values of $\lambda$, $\kappa/\lambda$,
and $\sin2\beta$ much like in Region~II; and the other with larger values 
of $\lambda$ and $\sin2\beta\sim \kappa/\lambda \sim 1$ similar
to Region~I.  In both allowed regions,
there exists an approximate $U(1)_R$ symmetry in the singlet sector
that is realized as a hierarchy between $A_{\lambda}$ and $A_{\kappa}$
relative to $\mu$.  The Higgs boson phenomenology within these
two regions is qualitatively the same as in Regions~I and II discussed
in Section~\ref{free}.

  Let us also mention that in the parameter scans, we do 
not demand that the lightest NMSSM superpartner be neutral.
For compactification scales below about $M_c \lesssim 10^{16}\,\gev$,
the lightest MSSM superpartner in minimal gaugino mediation
is often a mostly-right-handed stau~\cite{Kaplan:1999ac,Chacko:1999mi,
Schmaltz:2000gy}.  
The true LSP in this case is typically the 
gravitino~\cite{Kaplan:1999ac,Buchmuller:2005rt}.  
At larger values of the compactification scale, the lightest
MSSM superpartner in minimal gaugino mediation is usually
a mostly-Bino neutralino.  Within the NMSSM, there can also
arise a very light mostly-singlet neutralino state
when $\kappa/\lambda$ is small, as we discussed 
in Sections~\ref{free} and \ref{gauge}.  Such a state can supplant
the stau as the lightest NMSSM superpartner in gaugino mediation, 
providing another way to get around the problem of a charged LSP.  
This also leads to new possibilities for the production and 
identity of the dark matter, such as by the decoupling of a singlet 
neutralino LSP~\cite{Barger:2006kt}, or through the superWIMP 
scenario~\cite{Feng:2003uy} with a singlet neutralino NLSP.
This latter possibility, however, is likely constrained by the 
overproduction of hadronic debris after the onset of 
nucleosynthesis~\cite{Feng:2003uy,Kawasaki:2004qu}.

\subsection{Higgs Phenomenology with Vector-Like Exotics}

  A second modification of minimal gaugino mediation that can improve
the prospects for electroweak symmetry breaking in the NMSSM 
is to add vector-like exotics to the theory.  As in Section~\ref{gauge}
we consider $\ffb$ exotics with superpotential couplings
to the singlet given by Eq.~\eqref{newyuk}.
We again search for phenomenologically acceptable parameter 
regions by scanning over the model parameters.  
Our scans encompass the same ranges of input values 
as listed above, along with the exotic couplings
\beq
\xi_D \in [0,1],~~~\xi_L\in[0,1].
\eeq
When we include more than one set of exotics, we assume that 
the values of $\xi_D$ and $\xi_L$ are the same for all
exotic flavors.  Gaugino mediated boundary conditions are imposed 
at the compactification scale $M_c$ for all the soft terms, 
including $m_S^2$.  

  Our search strategy is similar to Section~\ref{gauge}, 
and consists of specifying $\lambda$, $\xi_D$, $\xi_L$, 
and $M_{1/2}$ at the compactification scale $M_c$, 
along with $\tan\beta$ near the electroweak scale,
and computing the low-energy spectrum that results from these input
parameters.  In doing so, we use a modified version of 
NMSPEC/NMHDECAY~\cite{Ellwanger:2004xm}
to perform the RG evolution of the model parameters and to find the low-scale 
values of $\kappa$, $|\mu|$, and $m_S^2$.  The value of $m_S^2$ computed
in this way will not usually agree with the gaugino mediation 
boundary condition of $m_S^2(M_c) \simeq 0$.  To correct for this,
we adjust the value of $\xi_D$ (or $\xi_L$), repeat the running, 
and iterate until the value of $|m_S^2|$ at the input scale 
$M_c$ lies below a specified small cutoff value.  Again, we demand 
that all couplings remain perturbative up to $M_c$, 
but we do not require the lightest NMSSM superpartner to be neutral.

  The results of our scans are very similar to what we found 
Section~\ref{gauge}, and illustrated in Fig.~\ref{gauge-ex-lk}.  
With a single set of $\ffb$ exotics, the phenomenologically 
allowed region of the parameter space is nearly identical to the small 
$\lambda$ region discussed in Section~\ref{gauge}
as well as Region~II studied in Section~\ref{free}. 
With two or more sets of $\ffb$ exotics, we find a second disjoint
allowed parameter region with $\lambda \sim \kappa > 0.4$
and $\tan\beta \lesssim 2.5$.  This region of the parameter space
is much the same as the large $\lambda$ region discussed in 
Section~\ref{gauge}.  In particular, the new exotic Yukawa
couplings $\xi_D$ and $\xi_L$ help to transmit the $U(1)_R$ breaking
from the gauginos to the singlet sector, leading to somewhat larger
values for the trilinear $A$ terms at the low scale.  The mostly-singlet
pseudoscalar $a_s$ is always heavier than the $h^0$ Higgs as a result,
leading to a Higgs collider phenomenology nearly identical to the MSSM
with a pseudoscalar mass in excess of $500\,\gev$.

\section{Conclusions\label{concl}}

  In the present work we have studied two simple deformations of
gauge and gaugino mediation within the NMSSM.  The first deformation
consists of allowing the singlet soft mass $m_S^2$ to be a free parameter
at the mediation scale, such as might arise if the singlet couples
to a $U(1)_R$ preserving source of supersymmetry breaking.  The second
deformation involves adding vector-like exotics with superpotential
couplings to the singlet superfield.  Both deformations
facilitate electroweak symmetry breaking by driving the singlet
soft mass $m_S^2$ to negative values in the infrared.  

  Near the electroweak scale, for either deformation of both minimal
gauge and gaugino mediation, we find a hierarchy between the value
of the effective $\mu$ parameter, $\mu = \mu_{eff} = \lambda\,v_s$,
and the $A_{\lambda}$ and $A_{\kappa}$ soft trilinear couplings,
as well as $\lambda\,v$.  The relative smallness of the singlet 
$A$ terms leads to an approximate $U(1)_R$ symmetry in the singlet sector, 
and consequently to a light pNGB pseudoscalar when this would-be 
symmetry is spontaneously broken.  The presence of this light 
pseudoscalar can have a large effect on the Higgs signatures of the theory.

  We find two distinct ways in which these deformations can
lead to consistent electroweak symmetry with a phenomenologically
acceptable Higgs boson spectrum.  The first and more interesting case
requires $|m_S^2/\mu^2|$ as well as $\kappa/\lambda$ and $\sin2\beta$
to be on the order of unity near the electroweak scale, 
along with $\lambda \gtrsim 0.4$.  With these parameter values,
there is an additional $F$-term contribution to the mass of the
lightest $CP$-even Higgs boson.  If, in addition, $A_{\kappa}$ and
$A_{\lambda}$ remain small near the electroweak scale, there is a 
SM-like Higgs boson state in the spectrum that can decay predominantly 
into pairs of the light pNGB $a_s$ pseudoscalar leading to new Higgs 
boson signatures at the LHC.  
These pseudoscalars usually decay primarily into $b\bar{b}$,
but can have a dominant branching into $\tau\bar{\tau}$
if they are particularly light ($m_{a_s} < 10\,\gev$).
This scenario is realized most easily 
in both gauge and gaugino mediation when $m_S^2$ is allowed to 
be free, with no additional contributions to the singlet 
trilinear couplings at the mediation scale.  

  The second way to obtain consistent electroweak symmetry breaking
within the deformations considered here is to have $\kappa/\lambda 
\sim \sin2\beta \ll 1$ near the electroweak scale.  This occurs
when $|m_S^2|/\mu^2 \ll 1$ with small singlet $A$ terms.  For the
electroweak symmetry breaking extremum to be stable, it is then
necessary to have $\lambda \ll 1$.  In this case, the singlet sector
couples only very weakly to the MSSM states, and therefore mostly decouples.  
It is still possible to have a very light pNGB pseudoscalar, but since
it interacts only feebly with the SM-like Higgs state in the spectrum,
the decays of this Higgs boson into pseudoscalar pairs are extremely rare.
In general, the Higgs phenomenology in this case is very similar to 
the MSSM with a relatively heavy pseudoscalar Higgs boson.
Despite the decoupling property of this region, the supersymmetric
phenomenology can still be modified if the lightest NMSSM superpartner
state is a mostly-singlet neutralino.

  Let us emphasize that we have only considered simple deformations
of \emph{minimal} gauge and gaugino mediation.  In these minimal versions,
we always find relatively large values for the effective $\mu$ parameter
near the electroweak scale.  By allowing for non-minimal versions
of these mediation mechanisms, it may be possible for much smaller
values of $\mu$ to emerge.  In a scenario with $\mu \sim \lambda\,v$,
there may be new ways to obtain a consistent pattern of electroweak
symmetry breaking, possibly also with decays of a SM-like Higgs bosons
into pairs of light pseudoscalars.
While not all models of supersymmetry breaking will give rise to modified
Higgs boson phenomenology, it is exciting that very minor modifications to
models as simple and well-motivated as gauge mediation and gaugino
mediation can.  This emphasiszes the need to search for Higgs bosons at
the LHC with as broad a net as possible.

\section*{Acknowledgements}

We thank Kaustaubh Agashe, Radovan Derm\'i\v sek, for frequent 
discussions in the early stages of this work.  
We also would like to acknowledge discussions with  Arjun Menon.  
The work of AP is supported by NSF CAREER Grant NSF-PHY-0743315.
The work of DM is supported by DOE Grant DE-FG02-95ER40899.


\appendix

\vspace{0.5cm}
\begin{flushleft}
\textbf{\Large{Appendix}}
\end{flushleft}
\vspace{-0.9cm}

\section{Higgs Boson Masses and Mixings\label{appa}}

  In this Appendix we present analytic expressions for
the Higgs boson mass eigenvalues and mixing matrices valid in the
limit of an approximate $U(1)_R$ symmetry in the singlet 
sector and $\lambda\,v \ll |\mu|$.  Our results extend the
findings of Ref.~\cite{Dobrescu:2000jt}.  A related expansion valid in 
the limit of an approximate $U(1)_{PQ}$ symmetry can be
found in Ref.~\cite{Miller:2003ay}.

  The $CP$-odd Higgs bosons of the NMSSM are made up of $Im(S)/\sqrt{2}$
and the non-Goldstone combination $A^0_v$ of $Im(H_u^0)\sqrt{2}$
and $Im(H_d^0)/\sqrt{2}$.  In the $\{Im(S)/\sqrt{2},\,A_v^0\}$ basis
the mass matrix is
\beq
\mathcal{M}_A^2 = \lrf{\mu}{\ltk}^2\,\left(
\begin{array}{cc}
\lrf{\stb}{2\ltk}\,(4+\alpha)\gamma^2-3\delta&\gamma(\alpha-2)\\
\cdot&\lrf{\stb}{2\ltk}^{-1}\,(1+\alpha)
\end{array}
\right).
\eeq
In writing this expression, we have defined the variables
\beq
\ltk = \frac{\lambda}{\kappa},~~~~~
\gamma = \frac{\lambda}{\kappa}\frac{\lambda\,v}{\mu},~~~
\alpha = \alfer,~~~
\delta = \frac{\lambda}{\kappa}\frac{A_{\kappa}}{\mu}.
\eeq
We will assume that $\gamma$, $\alpha$, and $\delta$ are all much
less than unity, and treat $\ltk$ and $\stb$ as being on the
order of unity.\footnote{This expansion also works quite well for small 
$\kappa/\lambda$ and $\stb$ up to higher-order terms in $\alpha$.} 
Under these assumptions, the $\mathcal{M}^2_{A_{22}}$ 
element is much larger than the $\mathcal{M}^2_{A_{11}}$ and 
$\mathcal{M}^2_{A_{12}}$ elements.
Thus, the rotation angle to diagonalize the mass matrix will
be small and the mass eigenstates will be close to $Im(S)/\sqrt{2}$
and $A_v^0$.  Labelling these mass eigenstates by $\{a_s,\,A^0\}$, 
we find the mixing matrix to be
\beq
\left(
\begin{array}{c}
a_s\\A^0
\end{array}
\right)
= \mathcal{O}
\left(
\begin{array}{c}
Im(S)/\sqrt{2}\\A_v^0
\end{array}
\right),
\eeq
with the approximate rotation matrix given by
\bea
\mathcal{O}_{11} &=& 
1-\frac{\stb^2}{2\ltk^2}(1-3\alpha)\gamma^2
= \mathcal{O}_{22} 
\\
\mathcal{O}_{12} &=& 
\frac{\stb}{2\ltk}\gamma\left(2- 3\alpha 
+ 3\alpha^2-3\frac{\stb}{\ltk}\delta\right)
-\frac{\stb^3}{2\ltk^3}\gamma^3
= -\mathcal{O}_{21}  .
\eea
In deriving these expressions we have treated 
$\alpha,\,\gamma = \mathcal{O}(\epsilon)$ 
and $\delta = \mathcal{O}(\epsilon^2)$ with $\epsilon \ll 1$, 
and we have kept terms only up to $\mathcal{O}(\epsilon^3)$. 

  The corresponding mass eigenvalues to this level of approximation are
\bea
m_{a_s}^2\,\left(\frac{\ltk}{\mu}\right)^2 &=&
-3\delta + {9}\alpha\frac{\stb}{2\ltk}\gamma^2,
\label{cpodd1a}\\
m_{A^0}^2\,\left(\frac{\ltk}{\mu}\right)^2 &=& 
\frac{2\ltk}{\stb}(1+\alpha)+ 4\frac{\stb}{2\ltk}(1-2\alpha)\gamma^2.
\label{cpodd2a}
\eea 

  To discuss the NMSSM $CP$-even Higgs mass matrices and their eigenvalues,
it is convenient to work in the basis $\{h_v^0,\,h_s^0,H_v^0\}$,
where $h_v^0$ is the combination of $Re(H_u^0)/\sqrt{2}$ and 
$Re(H_d^0)/\sqrt{2}$ that has the same tree-level couplings to 
the electroweak gauge bosons as the SM Higgs, $H_v^0$ is the orthogonal
combination, and $h_s^0 = Re(S)/\sqrt{2}$.  The transformation to this
basis is simply
\beq
\left(
\begin{array}{c}
Re(H_u^0)/\sqrt{2}\\
Re(H_d^0)/\sqrt{2}\\
Re(S)/\sqrt{2}
\end{array}
\right)
=
\left(
\begin{array}{ccc}
\cos\beta&\sin\beta&0\\
-\sin\beta&\cos\beta&0\\
0&0&1
\end{array}
\right)
\left(
\begin{array}{c}
H_v^0\\
h_v^0\\
h_s^0
\end{array}
\right).
\eeq

  The $CP$-even symmetric mass matrix in the 
basis $\{h_v^0,\,h_s^0,\,H_v^0\}$ reads
\beq
\mathcal{M}_H^2 = \lrf{\mu}{\ltk}^2\left(
\begin{array}{ccc}
\gamma^2(\stb^2+\tilde{g}^2\,\ctb^2)&2\ltk\gamma 
- \gamma(2+\alpha)\stb&\gamma^2(1-\tilde{g}^2)\stb\ctb\\
\cdot&4+\delta+\frac{\stb}{2\ltk}\alpha\gamma^2&-\gamma(2+\alpha)\ctb\\
\cdot&\cdot&(1+\alpha)\frac{2\ltk}{\stb} - \gamma^2(1-\tilde{g}^2)\stb^2
\end{array}
\right).
\eeq
Here, we have defined 
\beq
\tilde{g}^2 = \frac{(g^2+g'^2)}{2\lambda^2}.
\eeq
The $\{h_v^0,\,h_s^0,\,H_v^0\}$ basis is useful because all the
mixing elements in this matrix are suppressed by at least a factor
of $\gamma$, while the lower two diagonal elements are of order unity.
Since this mixing is small, we will designate the mass eigenvalues
by $\{h^0,\,h_s,\,H^0\}$ in analogy with the corresponding MSSM states.

  The transformation to the mass eigenbasis is
\beq
\left(
\begin{array}{c}
h^0\\h_s\\H^0
\end{array}
\right)
= U\,
\left(
\begin{array}{c}
h_v^0\\h_s^0\\H_v^0
\end{array}
\right),
\eeq
with the unitary matrix $U$ given by

\bea
U_{11} &=& 1 - \frac{\gamma^2}{32}\left[2\ltk-(2+\alpha)\stb\right]^2\\
\nnmb\\
U_{12} &=& \frac{\gamma}{16}\left((-4+\delta)[2\ltk-(2+\alpha)\stb]\right)\\
&& + \frac{\gamma^3}{16}\left( (\ltk-\stb)(3\ltk^2-6\ltk\stb+\stb^2)\right.
\nnmb\\
&&~~~\left.-2\frac{\ctb^2}{\ltk}[2\ltk\stb
+\tilde{g}^2(\ltk - 2\stb)(\ltk+\stb)]\right)
\nnmb\\
\nnmb\\
U_{21} &=& -\,\frac{\gamma}{16}\left((-4+\delta)
([2\ltk-(2+\alpha)\stb])\right)\\
&& - \frac{\gamma^3}{16}\left((\ltk-\stb)(3\ltk^2-6\ltk\stb+\stb^2)\right.
\nnmb\\
&&\left.-\frac{2\ctb^2}{(\ltk-2\stb)^2}\left[
2\stb(\ltk^2-3\ltk\stb+\stb^2)+\tilde{g}^2\ltk(\ltk-3\stb)(\ltk-2\stb)\right]
\right)
\nnmb
\\
\nnmb\\\nnmb\\
U_{22} &=& 1 -\frac{1}{8}\gamma^2\left((\ltk-\stb)[\ltk-(1+\alpha)\stb]
\right.\\
&&\left.-\frac{4\stb^2\ctb^2}{(\ltk-2\stb)^3}
[(-1+\alpha)\ltk+2(1+\alpha)\stb]\right)
\nnmb
\eea
\bea
U_{13} &=& \frac{\gamma^2}{4}\frac{\stb\ctb}{\ltk}\left[(-2+\alpha)\ltk
+2(\alpha+\tilde{g}^2-\alpha\,\tilde{g}^2)\stb\right]
\\
\nnmb\\
U_{31} &=& -\frac{\gamma^2}{2}\left(\frac{\ctb\stb^2}{\ltk(\ltk-2\stb)}
[\ltk+\tilde{g}^2\ltk -2\tilde{g}\stb]\right.\\
&& \left.-\alpha\frac{\stb^2\ctb}{\ltk(\ltk-2\stb)^2}
[(2+\tilde{g}^2)\ltk^2-4\tilde{g}^2\ltk\stb-4(1-\tilde{g}^2)\stb^2]\right)
\nnmb\\
%
U_{23} &=& \gamma\left(\frac{\ctb\stb}{2(\ltk-2\stb)^2}
[(2-\alpha)\ltk-(4+2\alpha-\delta)\stb]
+\alpha^2\frac{\stb\ctb\ltk}{2(\ltk-2\stb)^3}(\ltk+2\stb)\right)
\\
&& -\gamma^3\frac{1}{8(\ltk-2\stb)^2}\ctb\stb\left(
\ltk^4-2(4+\tilde{g}^2)\ltk^3\stb+(19+10\tilde{g}^2)\ltk^2\stb^2
-4(5+3\tilde{g}^2)\ltk\stb^3+12\stb^2\right)
\nnmb\\
%
U_{32} &=& -\gamma\left(\frac{\stb\ctb}{2(\ltk-2\stb)^2}
[(2-\alpha)\ltk-(4+2\alpha-\delta)\stb]
+\alpha^2\frac{\ctb\stb\ltk}{2(\ltk-2\stb)^3}(\ltk+2\stb)
\right)\\
&& +\gamma^3\frac{\ctb\stb^3}{2\ltk(\ltk-2\stb)^3}\left(
3\ctb^2\ltk-(\ltk-2\stb)[(1+\tilde{g}^2)\ltk^2-4\tilde{g}^2\ltk\stb
+2\tilde{g}^2\stb^2]\right)
\nnmb\\
%
%
U_{33} &=& 1+\gamma^2\frac{\stb^2\ctb^2}{2(\ltk-2\stb)^3}
\left[(-1+\alpha)\ltk+2(1+\alpha)\stb\right].
\eea

  The $CP$-even Higgs mass eigenvalues are
\bea
m_{h^0}^2\,\left(\frac{\ltk}{\mu}\right)^2 &=& 
\left[\tilde{g}^2\ctb^2-\ltk^2+(2+\alpha)\stb\ltk-\alpha\stb^2\right],
\label{cpeven1a}\\
m_{h_s}^2 \,\left(\frac{\ltk}{\mu}\right)^2 &=& 
4+\delta
\label{cpeven2a}\\
&& +\gamma^2\left(\ltk^2 + \frac{\alpha\stb}{2\ltk} - (2+\alpha)\ltk\stb
+(1+\alpha)\stb^2 - \frac{2\ctb^2\stb}{(\ltk-2\stb)^2}[\ltk-2(1+\alpha)\stb]
\right),
\nnmb\\
%
m_{H^0}^2\,\left(\frac{\ltk}{\mu}\right)^2 &=& 
\frac{2\ltk(1+\alpha)}{\stb}
\label{cpeven3a}\\
&&+\gamma^2\left((-1+\tilde{g}^2)\stb^2
+\frac{2\ctb^2\stb}{(\ltk-2\stb)^2}[\ltk-2(1+\alpha)\stb]
\right).
\nnmb
%
\eea

\section{RG Equations with Exotics\label{appb}}

  We collect here the modifications to the RG equations that
arise when multiple sets of vector-like exotics are added
to the NMSSM.  Our notation conventions follow NMHDECAY.
The additional exotics we consider consist of $\tilde{D}\oplus \tilde{D}^c$,
and $\tilde{L}\oplus \tilde{L}^c$, with $\tilde{D} = (3,1,-1/3)$ 
and $\tilde{L} = (1,2,-1/2)$.
Taken together, these exotics have the quantum numbers of 
a $5\oplus\bar{5}$ of $SU(5)$.  We include the trilinear exotic 
superpotential couplings 
\beq
W \supset \xi_{D_i}\,S\,\tilde{D}_i\,\tilde{D}^c_i 
+ \xi_{\tilde{L}_j}\,S\,\tilde{L}_j\,\tilde{L}^c_j.
\label{newyuk1}
\eeq
We also add the soft terms
\bea
-\mathscr{L}_{soft}&\subset& m_{D_i}^2|\tilde{D}_i|^2 
+ m_{D_i^c}^2|\tilde{D}_i^c|^2
+m_{\tilde{L}_j}^2|\tilde{L}_j|^2 + m_{L_j^c}^2|\tilde{L}_j^c|^2\\
&& + \xi_{D_i}A_{D_i}\,S\,\tilde{D}_i\,\tilde{D}^c_i 
+ \xi_{L_j}A_{L_j}\,S\,\tilde{L}_j\,\tilde{L}^c_j.
\nnmb
\eea

  The RG equations for the superpotential couplings are
\bea
(16\pi^2)\,\frac{d\ln\lambda}{dt} &=&
(N_w+2)\lambda^2 + 2\kappa^2 +N_c(\lambda_t^2+\lambda_b^2) +\lambda_{\tau}^2
\nnmb\\
&&-2\left[(Y_{H_u}^2+Y_{H_d}^2)g'^2+2C_2g^2\right]
\nnmb\\
&&+ N_c\sum_i\xi_{D_i}^2+N_w\sum_j\xi_{L_j}^2,
\\
(16\pi^2)\,\frac{d\ln\kappa}{dt} &=& 3N_w\lambda^2 + 6\kappa^2 
+ 3N_c\sum_i\xi_{D_i}^2+3N_w\sum_j\xi_{L_j}^2),
\\
(16\pi^2)\,\frac{d\ln\xi_{D_k}}{dt} &=& 
N_w\lambda^2+2\kappa^2-2\left[(Y_{D_k}^2+Y_{D_k^c}^2){g'}^2+2\,C_3g_3^2\right]
\nnmb\\
&&+2\xi_{D_k}^2+ N_c\sum_i\xi_{D_i}^2+N_w\sum_j\xi_{L_j}^2,
\\
(16\pi^2)\,\frac{d\ln\xi_{L_k}}{dt} &=& 
N_w\lambda^2+2\kappa^2-2\left[(Y_{L_k}^2+Y_{L_k^c}^2){g'}^2+2\,C_2g^2\right]
\nnmb\\
&&+2\xi_{L_k}^2+ N_c\sum_i\xi_{D_i}^2+N_w\sum_j\xi_{L_j}^2.
\eea
Here $N_c=3$ is the number of colors and $N_w = 2$ is the number
of ``weak'' colors.

  For the trilinear $A$ terms, we have
\bea
(16\pi^2)\,\frac{dA_{\lambda}}{dt} &=&
2(N_w+2)\lambda^2A_{\lambda}+4\kappa^2A_{\kappa}
+2N_c(\lambda_t^2A_t+\lambda_b^2A_b) + 2\lambda_{\tau}^2A_{\tau}
\nnmb\\
&&+4\left[(Y_{H_u}^2+Y_{H_d}^2)g'^2M_1+2C_2g^2M_2\right]
\nnmb\\
&&+ 2N_c\sum_i\xi_{D_i}^2A_{D_i}+2N_w\sum_j\xi_{L_j}^2A_{L_i},
\\
(16\pi^2)\,\frac{dA_{\kappa}}{dt} &=& 12\kappa^2A_{\kappa} 
+6N_w\lambda^2A_{\lambda} 
+ 6(N_c\sum_i\xi_{D_i}^2A_{D_i}+N_w\sum_j\xi_{L_j}^2A_{L_i}),
\\
(16\pi^2)\,\frac{dA_{D_k}}{dt} &=&
2N_w\lambda^2A_{\lambda} + 4\kappa^2A_{\kappa} 
+ 4\left[(Y_{D_k}^2+Y_{D_k^c}^2)g'^2M_1+2\,C_3g_3^2M_3\right]
\nnmb\\
&&+4\xi_{D_k}^2A_{D_k}+ 2N_c\sum_i\xi_{D_i}^2A_{D_i}
+2N_w\sum_j\xi_{L_j}^2A_{L_j},
\\
(16\pi^2)\,\frac{dA_{L_k}}{dt} &=&
2N_w\lambda^2A_{\lambda} + 4\kappa^2A_{\kappa} 
+ 4\left[(Y_{L_k}^2+Y_{L_k^c}^2)g'^2M_1+2\,C_2g^2M_2\right]
\nnmb\\
&&+4\xi_{L_k}^2A_{L_k}+ 2N_c\sum_i\xi_{D_i}^2A_{D_i}
+2N_w\sum_j\xi_{L_j}^2A_{L_j}.
\eea

  Finally, the modifications to the running of the scalar soft masses is
\bea
(16\pi)^2\frac{dm_S^2}{dt} &=&
2N_w\lambda^2(m_{H_u}^2+m_{H_d}^2 +m_S^2+A_{\lambda}^2)
+4\kappa^2(3m_S^2+A_{\kappa}^2)
\nnmb\\
&&+ 2N_c\sum_i\xi_{D_i}^2(m_{D_i}^2+m_{D_i^c}^2+m_S^2+A_{D_i}^2)
\nnmb\\
&&+2N_w\sum_j\xi_{L_j}^2(m_{L_j}^2+m_{L_j^c}^2+m_S^2+A_{L_j}^2),
\\
(16\pi)^2\frac{dm_{D_i}^2}{dt} &=& 
2\xi_{D_i}^2(m_{D_i}^2+m_{D_i^c}^2+m_S^2+A_{D_i}^2)\\
&&~~-8(Y_{D_i}^2g'^2M_1^2+C_3g_3^2M_3^2)\!+\!2Y_{D_i}\zeta,\nnmb
\\
(16\pi)^2\frac{dm_{L_j}^2}{dt} &=& 
2\xi_{L_j}^2(m_{L_j}^2+m_{L_j^c}^2+m_S^2+A_{L_j}^2)\\
&&~~-8(Y_{L_j}^2g'^2M_1^2+C_2g^2M_2^2)\!+\!2Y_{L_j}\zeta.\nnmb
\eea
The quantity $\zeta$ in these expressions is the hypercharge $D$-term
which vanishes in many simple models of gauge mediation~\cite{Giudice:1998bp}.



\end{document}